\documentclass[12pt]{article}
\usepackage{amsmath,amssymb,amsthm,amsxtra,overpic,bbm,bm,epsfig,ulem}
\usepackage{color}
\usepackage{cite}

\usepackage{hyperref}
\usepackage{url}

\renewcommand{\theequation}{\thesection.\arabic{equation}}

\def\thefootnote{\fnsymbol{footnote}}
\allowdisplaybreaks[4]

\begin{document}

\vspace{0.2cm}

\begin{center}
{\large\bf Confronting the seesaw mechanism with neutrino oscillations: \\
a general and explicit analytical bridge}
\end{center}

\vspace{0.2cm}

\begin{center}
{\bf Zhi-zhong Xing$^{1,3,4}$}
\footnote{xingzz@ihep.ac.cn},
{\bf Jing-yu Zhu$^{2}$}
\footnote{zhujingyu@impcas.ac.cn (corresponding author)}
\\
{$^{1}$Institute of High Energy Physics,
Chinese Academy of Sciences, Beijing 100049, China \\
$^{2}$Institute of Modern Physics, Chinese Academy of Sciences, Lanzhou 730000, China \\
$^{3}$School of Physical Sciences,
University of Chinese Academy of Sciences, Beijing 100049, China \\
$^{4}$Center of High Energy Physics, Peking University, Beijing 100871, China}
\end{center}

\vspace{2cm}
\begin{abstract}
With the help of a full Euler-like block parametrization of the flavor structure
for the canonical seesaw mechanism, we present the {\it first} general and explicit
analytical calculations of the two neutrino mass-squared differences, three flavor
mixing angles and the effective Dirac CP-violating phase responsible for the
primary behaviors of neutrino oscillations. Such model-independent results will
pave the way for testing the seesaw mechanism at low energies.
\end{abstract}

\vspace{2cm}

\newpage

\def\thefootnote{\arabic{footnote}}
\setcounter{footnote}{0}

\setcounter{equation}{0}
\section{Introduction}

It is widely accepted that the canonical seesaw mechanism~\cite{Minkowski:1977sc,
Yanagida:1979as,GellMann:1980vs,Glashow:1979nm,Mohapatra:1979ia} has stood out as
the most natural and most economical extension of the standard model (SM) of
particle physics to explain both the origin of tiny neutrino masses and the
cosmic baryon number asymmetry via leptogenesis~\cite{Fukugita:1986hr}. But this
elegant mechanism contains 18 unknown flavor parameters,
\begin{itemize}
\item     3 heavy Majorana neutrino masses $M^{}_j$ (for $j = 4, 5, 6$),

\item     9 active-sterile flavor mixing angles $\theta^{}_{ij}$ (for $i = 1, 2, 3$
and $j = 4, 5, 6$),

\item     6 independent CP-violating phases $\delta^{}_{ij}$ (for
$i = 1, 2, 3$ and $j = 4, 5, 6$),
\end{itemize}
in a complete Euler-like block parametrization of its flavor structure~\cite{Xing:2007zj,
Xing:2011ur}, making the experimental tests of its validity extremely difficult. In
this situation, to what extent can we {\it model-independently} probe or constrain the
seesaw parameter space with the help of our accurate determination of the two neutrino
mass-squared differences ($\Delta m^2_{21}$ and $\Delta m^2_{31}$ or $\Delta m^2_{32}$),
three active flavor mixing angles ($\theta^{}_{12}$, $\theta^{}_{13}$ and $\theta^{}_{23}$)
and the Dirac CP-violating phase ($\delta$) of the $3 \times 3$
Pontecorvo-Maki-Nakagawa-Sakata (PMNS) lepton flavor mixing matrix~\cite{Pontecorvo:1957cp,
Maki:1962mu,Pontecorvo:1967fh} in the next-generation long-baseline reactor and accelerator
neutrino oscillation experiments (e.g., JUNO~\cite{JUNO:2015zny, JUNO:2021vlw},
DUNE~\cite{DUNE:2015lol} and Hyper-Kamiokande~\cite{Hyper-KamiokandeProto-:2015xww})?

To answer this absolutely meaningful question, one needs to calculate all the 6 aforementioned
neutrino oscillation quantities in terms of the 18 {\it original} seesaw flavor parameters.
Such a general analytical approach has not been developed, mainly because the calculation
itself is formidably lengthy. The purpose of this work is to fill the gap, so as to pay
the way for testing the seesaw mechanism at low energies. Our strategies for doing so are
outlined below.

(1) We start from the seesaw-modified Lagrangian for leptonic weak charged-current
interactions:
\begin{align}
-{\cal L}^{}_{\rm cc} = & \hspace{0.1cm} \frac{g}{\sqrt{2}} \hspace{0.1cm}
\overline{\big(\begin{matrix} e & \mu & \tau\end{matrix}\big)^{}_{\rm L}}
\hspace{0.1cm} \gamma^\mu \left[ U \left( \begin{matrix} \nu^{}_{1}
\cr \nu^{}_{2} \cr \nu^{}_{3} \cr\end{matrix} \right)^{}_{\hspace{-0.08cm} \rm L}
+ R \left(\begin{matrix} N^{}_4 \cr N^{}_5 \cr N^{}_6
\cr\end{matrix}\right)^{}_{\hspace{-0.08cm} \rm L} \hspace{0.05cm} \right] W^-_\mu
+ {\rm h.c.} \; ,
\label{1.1}
\end{align}
where $U = A U^{}_0$ is the general PMNS matrix in which $U^{}_0$ denotes a unitary
matrix of the form
\begin{eqnarray}
U^{}_0 = \left( \begin{matrix} c^{}_{12} c^{}_{13} & \hat{s}^*_{12}
c^{}_{13} & \hat{s}^*_{13} \cr
-\hat{s}^{}_{12} c^{}_{23} -
c^{}_{12} \hat{s}^{}_{13} \hat{s}^*_{23} & c^{}_{12} c^{}_{23} -
\hat{s}^*_{12} \hat{s}^{}_{13} \hat{s}^*_{23} & c^{}_{13}
\hat{s}^*_{23} \cr
\hat{s}^{}_{12} \hat{s}^{}_{23} - c^{}_{12}
\hat{s}^{}_{13} c^{}_{23} & -c^{}_{12} \hat{s}^{}_{23} -
\hat{s}^*_{12} \hat{s}^{}_{13} c^{}_{23} & c^{}_{13} c^{}_{23}
\cr \end{matrix} \right) \;
\label{1.2}
\end{eqnarray}
with $c^{}_{12} \equiv \cos\theta^{}_{12}$ and $\hat{s}^{}_{12} \equiv
e^{{\rm i}\delta^{}_{12}} \sin\theta^{}_{12}$ and so on, $A$ is a
$3 \times 3$ lower-triangular matrix which deviates slightly from the identity
matrix $I$, and $R$ describes the strengths of Yukawa interactions of massive
Majorana neutrinos and satisfies the unitarity condition
$U U^\dagger + R R^\dagger = A A^\dagger + R R^\dagger = I$.
Given the Euler-like block parametrization of the seesaw flavor
structure proposed in Refs.~\cite{Xing:2007zj,Xing:2011ur}, $A$ and $R$
consist of the same set of active-sterile flavor mixing parameters (i.e.,
$\theta^{}_{ij}$ and $\delta^{}_{ij}$ for $i = 1, 2, 3$ and $j = 4, 5, 6$)
as can be seen in Eq.~(\ref{2.1}), and a proper redefinition of the
phases associated with three charged lepton fields in Eq.~(\ref{1.1})
allows us to remove three of the nine phase parameters $\delta^{}_{ij}$ or
their combinations and thus arrive at six independent CP-violating phases.

(2) We make use of the exact seesaw relation~\cite{Xing:2023adc}
\begin{align}
U^{}_0 D^{}_\nu U^T_0 = - \left(A^{-1} R\right) D^{}_N
\left(A^{-1} R\right)^T \;
\label{1.3}
\end{align}
to bridge the gap between the active and sterile neutrino sectors, where
$D^{}_\nu \equiv {\rm Diag}\{m^{}_1 , m^{}_2 , m^{}_3\}$
and $D^{}_N \equiv {\rm Diag}\{M^{}_4 , M^{}_5 , M^{}_6\}$ with $m^{}_i$
and $M^{}_j$ standing respectively for the light and heavy Majorana neutrino masses
(for $i = 1, 2, 3$ and $j = 4, 5, 6$). So it becomes clear that
the 9 flavor parameters of $D^{}_\nu$ and $U^{}_0$ (i.e.,
$m^{}_1$, $m^{}_2$, $m^{}_3$; $\theta^{}_{12}$, $\theta^{}_{13}$, $\theta^{}_{23}$;
$\delta^{}_{12}$, $\delta^{}_{13}$, $\delta^{}_{23}$) are all {\it derivational} in
the sense they can be derived from the 18 {\it original} seesaw flavor
parameters hidden in $D^{}_N$ and $R$. But these two sets of flavor parameters
are correlated with each other via Eq.~(\ref{1.3}) in a very complicated
way, so that the general analytical results for their explicit correlations
have never been achieved before.

(3) We only adopt the leading term of Maclaurin's series
\begin{eqnarray}
A^{-1} R =
\left(\begin{matrix} \hat{s}^*_{14} & \hat{s}^*_{15} & \hat{s}^*_{16} \cr
\hat{s}^*_{24} & \hat{s}^*_{25} & \hat{s}^*_{26} \cr
\hat{s}^*_{34} & \hat{s}^*_{35} & \hat{s}^*_{36} \cr \end{matrix}\right)
+ {\cal O}\left(s^3_{ij}\right) \; ,
\label{1.4}
\end{eqnarray}
where $i = 1, 2, 3$ and $j = 4, 5, 6$, so as to simplify our analytical
calculations and show how the light degrees of freedom depend on the original
seesaw flavor parameters in an instructive manner. This approximation is
very reliable, because every active-sterile flavor mixing angle in
$A^{-1}R$ is expected to be smaller than ${\cal O}(10^{-1.5}) \sim 3^\circ$,
as can be seen from Eqs.~(\ref{2.3}) and (\ref{2.4}) in section 2. It is
essentially equivalent to neglecting tiny non-unitarity of the PMNS matrix $U$,
or to focusing on the unique Weinberg operator in the seesaw-extended SM
effective field theory (SMEFT)~\cite{Weinberg:1979sa}.

(4) We determine the effective Dirac CP-violating phase of $U^{}_0$ (i.e.,
$\delta \equiv \delta^{}_{13} - \delta^{}_{12} - \delta^{}_{23}$ when taking the
standard phase convention of $U^{}_0$ as advocated by the Particle Data
Group~\cite{ParticleDataGroup:2024cfk}) with the help of its simple relation to
the Jarlskog invariant of leptonic CP violation~\cite{Jarlskog:1985ht,Wu:1985ea},
\begin{eqnarray}
{\cal J}^{}_\nu = \frac{1}{8} \sin 2\theta^{}_{12} \sin 2\theta^{}_{13}
\cos\theta^{}_{13} \sin 2\theta^{}_{23} \sin\delta \; ,
\label{1.5}
\end{eqnarray}
because the analytical expression of ${\cal J}^{}_\nu$ has recently been derived by
one of us in terms of the original seesaw flavor parameters and in the same
analytical approximation~\cite{Xing:2024xwb}.

(5) We leave aside the two effective Majorana CP-violating phases of $U^{}_0$
(i.e., $\rho \equiv \delta^{}_{12} + \delta^{}_{23}$ and $\sigma \equiv
\delta^{}_{23}$ in the PDG-like phase convention of $U$~\cite{Xing:2020ijf})
in this work for two reasons. First, they are irrelevant to the present and
future neutrino oscillation experiments. Second, they can be determined from
comparing the elements of $M^{}_\nu \equiv U^{}_0 D^{}_\nu U^T_0$ expressed in
terms of the light degrees of freedom with those obtained in terms of the original
seesaw flavor parameters~\cite{Xing:2023adc}, after the analytical results of
$m^{}_1$, $m^{}_2$, $m^{}_3$, $\theta^{}_{12}$, $\theta^{}_{13}$, $\theta^{}_{23}$
and $\delta$ are taken into account
\footnote{It is worth pointing out that the effective mass
$\langle m\rangle^{}_{0\nu2\beta}$ for a neutrinoless double-beta decay depends
on both $m^{}_i U^2_{e i}$ (for $i = 1, 2, 3$) and $R^2_{e j}/M^{}_j$ (for
$j = 4, 5, 6$) in the canonical seesaw mechanism~\cite{Xing:2020ijf}, as
both light and heavy Majorana neutrinos contribute to the same
lepton-number-violating process of this kind. As a result,
$\langle m\rangle^{}_{0\nu2\beta}$ is simply a function of the original seesaw
flavor parameters $M^{}_j$ and $R^{}_{e j}$ thanks to the exact seesaw
relation $\left(m^{}_1 U^2_{e 1} + m^{}_2 U^2_{e 2} + m^{}_3 U^2_{e 3}\right)
= -\left(M^{}_4 R^2_{e 4} + M^{}_5 R^2_{e 5} + M^{}_6 R^2_{e 6}\right)$ as
indicated by Eq.~(\ref{1.3}).}.

The remaining parts of this paper are organized as follows. In section 2 we show
our numerical constraints on the 9 active-sterile flavor mixing angles to assure
the validity and reliability of our analytical approximations made in this work.
Then we calculate the light neutrino masses, the active flavor mixing angles
and the effective Dirac CP-violating phase in section 3. Section 4 is devoted
to the simplified results in the {\it minimal} seesaw scenario with only two
right-handed neutrino fields, and section 5 is devoted to our summary and
some further discussions.

\setcounter{equation}{0}
\section{Bounds on active-sterile flavor mixing}
\label{section 2}

Taking account of a full Euler-like block parametrization of the flavor
structure for the canonical seesaw mechanism~\cite{Xing:2007zj,Xing:2011ur},
we write out the explicit expressions of $A$ and $R$ appearing in Eq.~(\ref{1.1}):
\begin{eqnarray}
A \hspace{-0.2cm} & = & \hspace{-0.2cm}
\left( \begin{matrix} c^{}_{14} c^{}_{15} c^{}_{16} & 0 & 0
\cr \vspace{-0.45cm} \cr
\begin{array}{l} -c^{}_{14} c^{}_{15} \hat{s}^{}_{16} \hat{s}^*_{26} -
c^{}_{14} \hat{s}^{}_{15} \hat{s}^*_{25} c^{}_{26} \\
-\hat{s}^{}_{14} \hat{s}^*_{24} c^{}_{25} c^{}_{26} \end{array} &
c^{}_{24} c^{}_{25} c^{}_{26} & 0 \cr \vspace{-0.45cm} \cr
\begin{array}{l} -c^{}_{14} c^{}_{15} \hat{s}^{}_{16} c^{}_{26} \hat{s}^*_{36}
+ c^{}_{14} \hat{s}^{}_{15} \hat{s}^*_{25} \hat{s}^{}_{26} \hat{s}^*_{36} \\
- c^{}_{14} \hat{s}^{}_{15} c^{}_{25} \hat{s}^*_{35} c^{}_{36} +
\hat{s}^{}_{14} \hat{s}^*_{24} c^{}_{25} \hat{s}^{}_{26}
\hat{s}^*_{36} \\
+ \hat{s}^{}_{14} \hat{s}^*_{24} \hat{s}^{}_{25} \hat{s}^*_{35}
c^{}_{36} - \hat{s}^{}_{14} c^{}_{24} \hat{s}^*_{34} c^{}_{35}
c^{}_{36} \end{array} &
\begin{array}{l} -c^{}_{24} c^{}_{25} \hat{s}^{}_{26} \hat{s}^*_{36} -
c^{}_{24} \hat{s}^{}_{25} \hat{s}^*_{35} c^{}_{36} \\
-\hat{s}^{}_{24} \hat{s}^*_{34} c^{}_{35} c^{}_{36} \end{array} &
c^{}_{34} c^{}_{35} c^{}_{36} \cr \end{matrix} \right) \; ,
\nonumber \\ \vspace{-0.5cm} \nonumber \\
R \hspace{-0.2cm} & = & \hspace{-0.2cm}
\left( \begin{matrix} \hat{s}^*_{14} c^{}_{15} c^{}_{16} &
\hat{s}^*_{15} c^{}_{16} & \hat{s}^*_{16} \cr \vspace{-0.45cm} \cr
\begin{array}{l} -\hat{s}^*_{14} c^{}_{15} \hat{s}^{}_{16} \hat{s}^*_{26} -
\hat{s}^*_{14} \hat{s}^{}_{15} \hat{s}^*_{25} c^{}_{26} \\
+ c^{}_{14} \hat{s}^*_{24} c^{}_{25} c^{}_{26} \end{array} & -
\hat{s}^*_{15} \hat{s}^{}_{16} \hat{s}^*_{26} + c^{}_{15}
\hat{s}^*_{25} c^{}_{26} & c^{}_{16} \hat{s}^*_{26} \cr \vspace{-0.45cm} \cr
\begin{array}{l} -\hat{s}^*_{14} c^{}_{15} \hat{s}^{}_{16} c^{}_{26}
\hat{s}^*_{36} + \hat{s}^*_{14} \hat{s}^{}_{15} \hat{s}^*_{25}
\hat{s}^{}_{26} \hat{s}^*_{36} \\ - \hat{s}^*_{14} \hat{s}^{}_{15}
c^{}_{25} \hat{s}^*_{35} c^{}_{36} - c^{}_{14} \hat{s}^*_{24}
c^{}_{25} \hat{s}^{}_{26}
\hat{s}^*_{36} \\
- c^{}_{14} \hat{s}^*_{24} \hat{s}^{}_{25} \hat{s}^*_{35}
c^{}_{36} + c^{}_{14} c^{}_{24} \hat{s}^*_{34} c^{}_{35} c^{}_{36}
\end{array} &
\begin{array}{l} -\hat{s}^*_{15} \hat{s}^{}_{16} c^{}_{26} \hat{s}^*_{36}
- c^{}_{15} \hat{s}^*_{25} \hat{s}^{}_{26} \hat{s}^*_{36} \\
+c^{}_{15} c^{}_{25} \hat{s}^*_{35} c^{}_{36} \end{array} &
c^{}_{16} c^{}_{26} \hat{s}^*_{36} \cr \end{matrix} \right) \; , \hspace{0.8cm}
\label{2.1}
\end{eqnarray}
where $c^{}_{ij} \equiv \cos\theta^{}_{ij}$ and $\hat{s}^{}_{ij} \equiv
e^{{\rm i}\delta^{}_{ij}} \sin\theta^{}_{ij}$ with $\theta^{}_{ij}$ and
$\delta^{}_{ij}$ being the active-sterile flavor mixing angles and their
associated CP-violating phases (for $i = 1, 2, 3$ and $j = 4, 5, 6$).
We require that all $\theta^{}_{ij}$ lie in the first quadrant, while
$\delta^{}_{ij}$ may vary between zero and $2\pi$.
Now that $A$ characterizes a deviation of the PMNS matrix $U = A U^{}_0$
from the unitary matrix $U^{}_0$, it must not be far away from
the identity matrix $I$. A careful global analysis of current electroweak
precision measurements and neutrino oscillation data has given the numerical
bounds~\cite{Antusch:2014woa,Blennow:2016jkn,Wang:2021rsi,Blennow:2023mqx}
\begin{eqnarray}
\epsilon \equiv \left| I - A\right| < \left\{
\begin{array}{l}
\left( \begin{matrix}
1.3 \times 10^{-3} & 0 & 0 \cr
2.4 \times 10^{-5} & 1.1 \times 10^{-5} & 0 \cr
1.8 \times 10^{-3} & 1.1 \times 10^{-4} & 1.0 \times 10^{-3} \cr \end{matrix}
\right) \;  \hspace{0.6cm} ({\rm NMO}) \;
\\ \vspace{-0.45cm} \\
\left( \begin{matrix}
1.4 \times 10^{-3} & 0 & 0 \cr
2.4 \times 10^{-5} & 1.0 \times 10^{-5} & 0 \cr
1.6 \times 10^{-3} & 3.6 \times 10^{-5} & 8.1 \times 10^{-4} \cr \end{matrix}
\right) \;  \hspace{0.6cm} ({\rm IMO}) \;
\end{array} \right.
\label{2.2}
\end{eqnarray}
in the normal mass ordering (NMO) or inverted mass ordering (IMO) case for
the three light Majorana neutrinos, implying that all
the 9 active-sterile flavor mixing angles are very small.
\begin{figure}[t]
\begin{center}
\includegraphics[width=13.35cm]{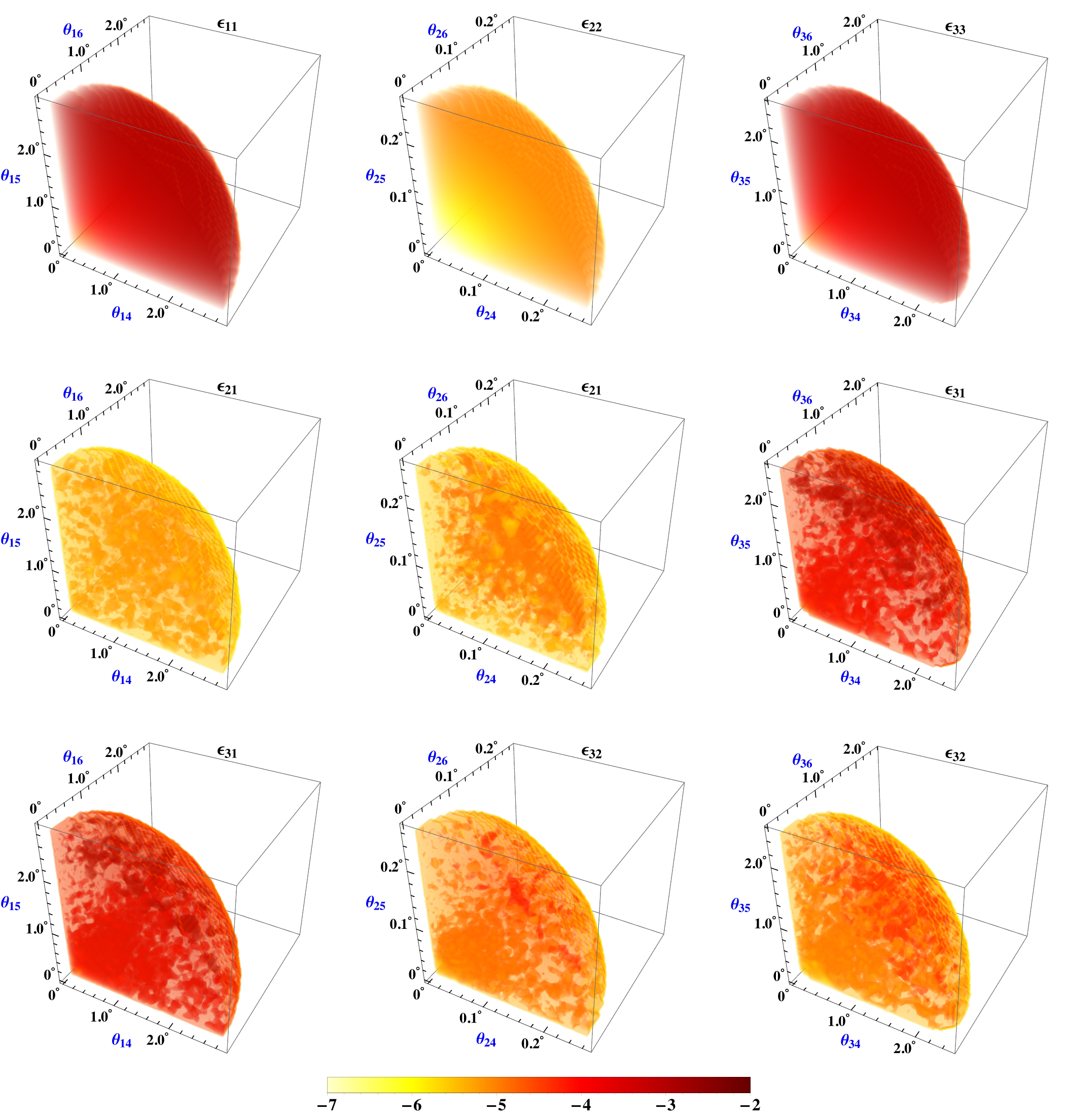}
\caption{Numerical constraints on the active-sterile flavor mixing angles
$\theta^{}_{1j}$, $\theta^{}_{2j}$ and $\theta^{}_{3j}$ (for $j = 4, 5, 6$)
with the help of the upper bounds on $\epsilon^{}_{ii^\prime}$ (for
$i, i^\prime = 1, 2, 3$) in Eq.~(\ref{2.2}) in the NMO case, where
different colors in the bar legend denote different values of
$\log_{10}\left(\epsilon^{}_{i i^{\prime}}\right)$.}
\label{fig1}
\end{center}
\end{figure}
\begin{figure}[t]
\begin{center}
\includegraphics[width=13.35cm]{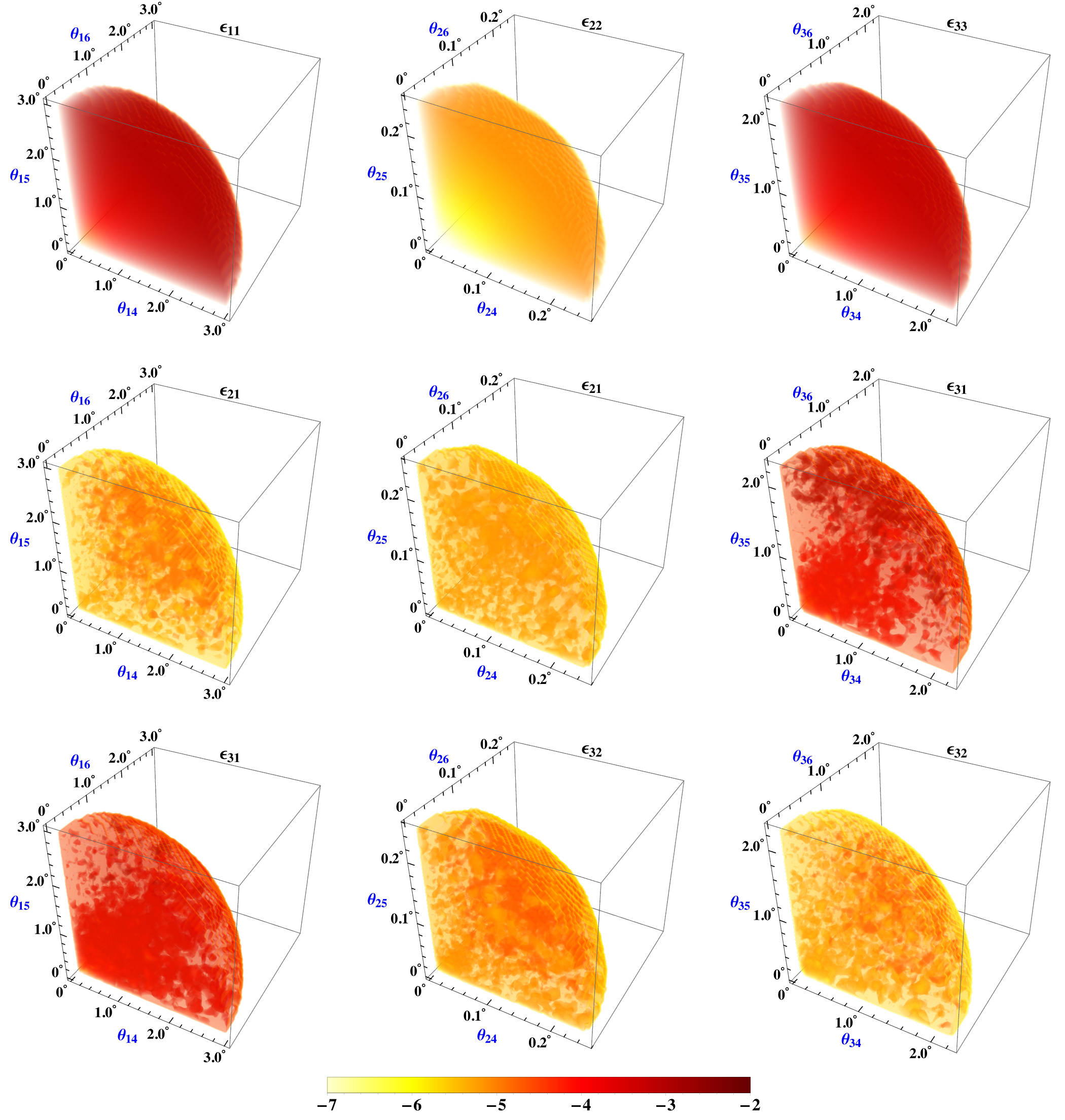}
\caption{Numerical constraints on the active-sterile flavor mixing angles
$\theta^{}_{1j}$, $\theta^{}_{2j}$ and $\theta^{}_{3j}$ (for $j = 4, 5, 6$)
with the help of the upper bounds on $\epsilon^{}_{ii^\prime}$ (for
$i, i^\prime = 1, 2, 3$) in Eq.~(\ref{2.2}) in the IMO case, where
different colors in the bar legend denote different values of
$\log_{10}\left(\epsilon^{}_{i i^{\prime}}\right)$.}
\label{fig2}
\end{center}
\end{figure}

In Figs.~\ref{fig1} and \ref{fig2}, we illustrate how the six nonzero elements
of $\epsilon$ depend on $\theta_{i4}^{}$, $\theta_{i5}^{}$ and $\theta_{i6}^{}$
(for $i=1,2,3$) by confronting Eq.~(\ref{2.1}) with Eq.~(\ref{2.2}) and taking
$\delta^{}_{ij} \in \left[0, 2\pi\right]$ in the respective NMO and IMO cases.
The upper bounds of $\theta^{}_{1j}$, $\theta^{}_{2j}$ and $\theta^{}_{3j}$
(for $j = 4, 5, 6$) can therefore be extracted as follows:
\begin{eqnarray}
\theta^{}_{1j} < 2.92^\circ \; , \quad
\theta^{}_{2j} < 0.27^\circ \; , \quad
\theta^{}_{3j} < 2.56^\circ \;
\label{2.3}
\end{eqnarray}
in the NMO case (i.e., $m^{}_1 < m^{}_2 < m^{}_3$);
or
\begin{eqnarray}
\theta^{}_{1j} < 3.03^\circ \; , \quad
\theta^{}_{2j} < 0.26^\circ \; , \quad
\theta^{}_{3j} < 2.31^\circ \;
\label{2.4}
\end{eqnarray}
in the IMO case (i.e., $m^{}_3 < m^{}_1 < m^{}_2$). These novel constraints
allow us to expand the elements of $A$ and $R$ as Maclaurin's series of the form
\begin{eqnarray}
A \hspace{-0.2cm} & = & \hspace{-0.2cm}
I - \frac{1}{2} \left( \begin{matrix} s^2_{14} +
s^2_{15} + s^2_{16} & 0 & 0 \cr \vspace{-0.4cm} \cr
2\hat{s}^{}_{14}
\hat{s}^*_{24} + 2\hat{s}^{}_{15} \hat{s}^*_{25} + 2\hat{s}^{}_{16}
\hat{s}^*_{26} & s^2_{24} + s^2_{25} + s^2_{26}
& 0 \cr \vspace{-0.4cm} \cr
2\hat{s}^{}_{14} \hat{s}^*_{34} + 2\hat{s}^{}_{15}
\hat{s}^*_{35} + 2\hat{s}^{}_{16} \hat{s}^*_{36} & 2\hat{s}^{}_{24}
\hat{s}^*_{34} + 2\hat{s}^{}_{25} \hat{s}^*_{35} + 2\hat{s}^{}_{26}
\hat{s}^*_{36} & s^2_{34} + s^2_{35} + s^2_{36} \cr \end{matrix}
\right) + {\cal O}\left(s^4_{ij}\right) \; ,
\nonumber \\
R \hspace{-0.2cm} & = & \hspace{-0.2cm}
\left( \begin{matrix} \hat{s}^*_{14} &
\hat{s}^*_{15} & \hat{s}^*_{16} \cr \vspace{-0.4cm} \cr
\hat{s}^*_{24} & \hat{s}^*_{25} &
\hat{s}^*_{26} \cr \vspace{-0.4cm} \cr
\hat{s}^*_{34} & \hat{s}^*_{35} &
\hat{s}^*_{36} \cr \end{matrix} \right) + {\cal O}\left(s^3_{ij}\right) \; .
\label{2.5}
\end{eqnarray}
So it is really reliable to use the leading term of $A^{-1} R$ shown by
Eq.~(\ref{1.4}) in our analytical calculations.

\setcounter{equation}{0}
\section{Results for the canonical seesaw mechanism}

\subsection{The masses of three light neutrinos}

To derive the expressions of $m_i^2$ (for $i=1, 2, 3$) in terms of the original
seesaw flavor parameters, let us begin from Eq. (\ref{1.3}) by defining
\begin{eqnarray}
&& H \equiv \left(U^{}_0 D^{}_\nu U^T_0\right)
\left(U^{}_0 D^{}_\nu U^T_0\right)^\dagger
= U_0^{} D_{\nu}^2 U_0^{\dagger} = R D_{N}^{} R^{T}_{} R^{*}_{} D_{N}^{}
R_{}^{\dagger} \; ,
\nonumber \\
&& H^2 \equiv H H^\dagger
= U_0^{} D_{\nu}^4 U_0^{\dagger} = R D_{N}^{} R^{T}_{} R^{*}_{} D_{N}^{}
R_{}^{\dagger}R D_{N}^{} R^{T}_{} R^{*}_{} D_{N}^{} R_{}^{\dagger} \; ,
\hspace{0.8cm}
\label{3.1}
\end{eqnarray}
where the reliable approximations made in Eqs.~(\ref{1.4}) and (\ref{2.5})
have been taken into consideration (i.e., only the leading terms of $A$ and $R$
are kept in our subsequent calculations).
The elements of hermitian $H$ and $H^2$ can be more explicitly expressed as
\begin{eqnarray}
&& H_{\alpha \beta}^{} = \sum^6_{j=4} \sum^6_{k=4}
M_j^{} M_k^{} I_{jk}^{} R_{\alpha j}^{} R_{\beta k}^* \; ,
\nonumber\\
&& \left(H^2\right)_{\alpha \beta}^{} =
\sum^6_{j=4} \sum^6_{k=4} \sum^6_{m=4} \sum^6_{n=4} M_j^{} M_k^{} M_m^{}
M_{n}^{} I_{jk}^{} I_{mn}^{} I_{mk}^{} R_{\alpha j}^{} R_{\beta n}^{*} \; ,
\hspace{0.9cm}
\label{3.2}
\end{eqnarray}
where the Greek subscripts run over $e$, $\mu$ and $\tau$, and
\begin{eqnarray}
I_{jk}^{} \equiv R_{e j}^{} R^{*}_{e k} + R_{\mu j}^{} R^{*}_{\mu k} +
R_{\tau j}^{} R^{*}_{\tau k} = \sum_{i=1}^3 \hat s_{ij}^{*} \hat s_{i k} \;
\label{3.3}
\end{eqnarray}
for $j, k = 4, 5, 6$. As ${\rm tr}\left(H\right) = m_1^2 + m_2^2 + m_3^2$,
${\rm tr}\left(H^2\right) = m_1^4 + m_2^4 +  m_3^4$ and
$\det\left(H\right) = m_1^2 m_2^2 m_3^2$ hold, $m^2_i$ (for $i = 1, 2, 3$)
must be the three roots of the cubic characteristic equation
\begin{eqnarray}
m^6_i - {\rm tr}\left(H\right) m^4_i +
\frac{\left[{\rm tr}\left(H\right)\right]^2 -
{\rm tr}\left(H^2\right)}{2} m^2_i - \det\left(H\right) = 0 \; .
\label{3.4}
\end{eqnarray}
Then we arrive at the solutions
\begin{eqnarray}
&& m_1^2 = \frac13 x - \frac13 \sqrt{x^2 -3 y}\left[z +
\sqrt{3 \left(1-z^2\right)}\right] \; , \hspace{1.6cm}
\nonumber \\
&& m_2^2 = \frac13 x - \frac13 \sqrt{x^2 -3 y}\left[z -
\sqrt{3 \left(1-z^2\right)}\right] \; ,
\nonumber \\
&& m_3^2 = \frac13 x + \frac23 z\sqrt{x^2 -3 y} \;
\label{3.5}
\end{eqnarray}
in the NMO case; or
\begin{eqnarray}
&& m_1^2 = \frac13 x - \frac13 \sqrt{x^2 -3 y}\left[z -
\sqrt{3 \left(1-z^2\right)}\right] \; , \hspace{1.6cm}
\nonumber \\
&& m_2^2 = \frac13 x + \frac23 z\sqrt{x^2 -3 y} \; ,
\nonumber \\
&& m_3^2 = \frac13 x - \frac13 \sqrt{x^2 -3 y}\left[z +
\sqrt{3 \left(1-z^2\right)}\right] \;
\label{3.6}
\end{eqnarray}
in the IMO case, where
\begin{eqnarray}
&& x = {\rm tr}\left(H\right) \; ,
\nonumber \\
&& y = \frac{1}{2} \Big[\left[{\rm tr}\left(H\right)\right]^2 -
{\rm tr}\left(H^2\right)\Big] \; ,
\nonumber \\
&& z = \cos \left[\frac13 \arccos \frac{2 x^3 - 9 x y + 27d}
{2 \left(x^2 - 3 y\right)^{3/2}} \right] \;  \hspace{1.9cm}
\label{3.7}
\end{eqnarray}
with $d = \det\left(H\right)$. The three neutrino mass-squared differences
$\Delta_{ii^{\prime}} \equiv m_i^{2} - m_{i^{\prime}}^{2}$ (for
$i \neq i^{\prime} = 1, 2, 3$) turn out to be
\begin{eqnarray}
\Delta_{21}^{} \hspace{-0.25cm} & = & \hspace{-0.25cm}
\frac{2}{3} \sqrt{x^2-3 y} \sqrt{3\left(1 - z^2\right)} \;\; ,
\nonumber \\
\Delta_{31}^{} \hspace{-0.25cm} & = & \hspace{-0.25cm}
\frac13 \sqrt{x^2 -3 y}\left[3z + \sqrt{3\left(1-z^2\right)}\right] \; ,
\hspace{1.2cm}
\nonumber \\
\Delta_{32}^{} \hspace{-0.25cm} & = & \hspace{-0.25cm}
\frac13 \sqrt{x^2 -3 y}\left[3z - \sqrt{3\left(1-z^2\right)}\right] \;
\label{3.8}
\end{eqnarray}
in the NMO case; or
\begin{eqnarray}
\Delta_{21}^{} \hspace{-0.25cm} & = & \hspace{-0.25cm}
\frac13 \sqrt{x^2 -3 y}\left[3z - \sqrt{3\left(1-z^2\right)}\right] \; ,
\hspace{1.2cm}
\nonumber \\
\Delta_{13}^{} \hspace{-0.25cm} & = & \hspace{-0.25cm}
\frac{2}{3} \sqrt{x^2-3 y} \sqrt{3\left(1 - z^2\right)} \;\; ,
\nonumber \\
\Delta_{23}^{} \hspace{-0.25cm} & = & \hspace{-0.25cm}
\frac13 \sqrt{x^2 -3 y}\left[3z + \sqrt{3\left(1-z^2\right)}\right] \;
\label{3.9}
\end{eqnarray}
in the IMO case. Taking account of Eq.~(\ref{3.2}), we have
\begin{eqnarray}
&& d = M_4^2 M_5^2 M_6^2 \left(I_{56}^{} I_{65}^{} I_{44}^{}
+ I_{46}^{} I_{64}^{} I_{55}^{}
+ I_{45}^{} I_{54}^{} I_{66}^{}
- I_{45}^{} I_{56}^{} I_{64}^{}
- I_{54}^{} I_{65}^{} I_{46}^{}
- I_{44}^{} I_{55}^{} I_{66}^{}\right)^2_{} \; ,
\nonumber \\
&& x = M_4^2 I_{44}^2 + M_5^2 I_{55}^2 + M_6^2 I_{66}^2 + M_4^{} M_5^{} \left(I_{45}^2
+ I_{54}^2\right) + M_4^{} M_6^{} \left(I_{46}^2 + I_{64}^2\right) +
M_5^{} M_6^{} \left(I_{56}^2 + I_{65}^2\right) \; , \hspace{1cm}
\nonumber \\
&& y = M_4^2 M_5^2 \left(I_{45}^{} I_{54}^{} - I_{44}^{} I_{55}^{}\right)^2 +
M_4^2 M_6^2 \left(I_{46}^{} I_{64}^{} - I_{44}^{} I_{66}^{}\right)^2 +
M_5^2 M_6^2 \left(I_{56}^{} I_{65}^{} - I_{55}^{} I_{66}^{}\right)^2
\nonumber \\
&& \hspace{0.8cm}
+ \left[\left(I_{54}^{} I_{46}^{} - I_{44}^{} I_{56}^{}\right)^2 +
\left(I_{45}^{} I_{64}^{} - I_{44} I_{65}^{}\right)^2\right]
M_4^2 M_5^{} M_6^{}
\nonumber \\
&& \hspace{0.8cm}
+ \left[\left(I_{45}^{} I_{56}^{} - I_{55} I_{46}^{}\right)^2 +
\left(I_{54}^{} I_{65}^{} - I_{55}^{} I_{64}^{}\right)^2 \right]
M_4^{} M_5^{2} M_6^{}
\nonumber \\
&& \hspace{0.8cm}
+ \left[\left(I_{56}^{} I_{64}^{} - I_{66}^{} I_{54}^{}\right)^2 +
\left(I_{65}^{} I_{46}^{} - I_{66}^{} I_{45}^{}\right)^2\right]
M_4^{} M_5^{} M_6^{2} \; ,
\label{3.10}
\end{eqnarray}
where $I^{}_{jk} = I^*_{kj}$ (for $j, k = 4, 5, 6$) have been defined in
Eq.~(\ref{3.3}).

To see how $x$, $y$ and $d$ depend on the active-sterile flavor mixing angles
and CP-violating phases, we proceed to rewrite their expressions in a more
explicit manner:
\begin{eqnarray}
d \hspace{-0.25cm} & = & \hspace{-0.25cm}
M_4^2 M_5^{2}M_6^{2} \Bigg[ I^{}_{44} I^{}_{55} I^{}_{66} - I_{44}^{}
\sum^3_{i=1} \sum_{i^{\prime} = 1}^3 s_{i 5}^{} s_{i 6}^{}
s_{i^{\prime} 5}^{} s_{i^{\prime} 6}^{} \cos\left(\beta_{i}^{} -
\beta_{i^{\prime}_{}}^{}\right)
- I_{55}^{} \sum^3_{i=1} \sum_{i^{\prime} = 1}^3
s_{i 4}^{} s_{i 6}^{} s_{i^{\prime} 4}^{} s_{i^{\prime} 6}^{}
\cos\left(\gamma_{i}^{} - \gamma_{i^{\prime}_{}}^{}\right)
\nonumber \\
\hspace{-0.25cm} && \hspace{-0.25cm}
- \hspace{0.06cm} I_{66}^{} \sum^3_{i=1} \sum_{i^{\prime} = 1}^3
s_{i 4}^{} s_{i 5}^{} s_{i^{\prime} 4}^{} s_{i^{\prime} 5}^{}
\cos\left(\alpha_{i}^{} - \alpha_{i^{\prime}_{}}^{}\right)
+ 2 \sum^3_{i=1} \sum_{i^{\prime} = 1}^3 \sum^3_{i^{\prime \prime}=1}
s_{i4}^{} s_{i5}^{} s_{i^{\prime} 5}^{} s_{i^{\prime} 6}^{}
s_{i^{\prime \prime} 4}^{} s_{i^{\prime\prime} 6}^{}
\cos\left(\alpha_{i}^{} + \beta_{i^{\prime}}^{} +
\gamma_{i^{\prime\prime}}^{}\right) \Bigg]^2 \; ,
\nonumber \\
x \hspace{-0.25cm} & = & \hspace{-0.25cm}
M_4^2 I_{44}^{2} + M_5^2 I_{55}^{2} + M_6^2 I_{66}^{2} + 2 M_4^{} M_5^{}
\sum^3_{i=1} \sum_{i^{\prime} = 1}^3
s_{i4}^{} s_{i5}^{} s_{i^{\prime} 4}^{} s_{i^{\prime} 5}^{}
\cos\left(\alpha_i^{} + \alpha_{i^{\prime}_{}}^{}\right)
\nonumber \\
\hspace{-0.25cm} && \hspace{-0.25cm}
+ \hspace{0.06cm} 2 M_5^{} M_6^{} \sum^3_{i=1} \sum_{i^{\prime} = 1}^3
s_{i5}^{} s_{i6}^{} s_{i^{\prime} 5}^{} s_{i^{\prime} 6}^{} \cos\left(\beta_i^{} +
\beta_{i^{\prime}_{}}^{}\right)
+ 2 M_4^{} M_6^{} \sum^3_{i=1} \sum_{i^{\prime} = 1}^3
s_{i4}^{} s_{i6}^{} s_{i^{\prime} 4}^{} s_{i^{\prime} 6}^{} \cos\left(\gamma_i^{} +
\gamma_{i^{\prime}_{}}^{}\right) \; , \hspace{1.4cm}
\nonumber \\
y \hspace{-0.25cm} & = & \hspace{-0.25cm}
2 M_4^2 M_5^{} M_6^{} \left\{ 2 \left[
\sum^3_{i=1} \sum_{i^{\prime} = 1}^3 s_{i4}^{} s_{i5}^{}
s_{i^{\prime}4}^{} s_{i^{\prime}6}^{} \cos\left(\alpha_i^{} +
\gamma_{i^\prime_{}}^{}\right) \right]^2 \right.
\nonumber \\
\hspace{-0.25cm} && \hspace{-0.25cm}
\left. - \left[ \sum^3_{i=1} \sum_{i^{\prime} = 1}^3
s_{i4}^{} s_{i5}^{} s_{i^{\prime}_{}4} s_{i^{\prime} 5}
\cos\left(\alpha_i^{} - \alpha_{i^{\prime}_{}}^{}\right)\right] \times
\left[ \sum^3_{i=1} \sum_{i^{\prime} = 1}^3 s_{i4}^{} s_{i6}^{}
s_{i^{\prime}_{}4} s_{i^{\prime} 6} \cos\left(\gamma_i^{} -
\gamma_{i^{\prime}_{}}^{}\right) \right] \right.
\nonumber \\
\hspace{-0.25cm} && \hspace{-0.25cm}
\left. - \hspace{0.06cm} 2 I_{44}^{}
\sum^3_{i=1} \sum_{i^{\prime} = 1}^3 \sum^3_{i^{\prime\prime}=1}
s_{i4}^{} s_{i5}^{} s_{ i^{\prime}_{} 5}^{} s_{ i^{\prime}_{} 6}^{}
s_{ i^{\prime\prime}_{} 4}^{} s_{i^{\prime\prime}_{} 6}^{}
\cos\left(\alpha_i^{} - \beta_{i_{}^{\prime}} + \gamma_{i_{}^{\prime\prime}}\right)
+ I_{44}^2 \sum^3_{i=1} \sum_{i^{\prime} = 1}^3
s_{i5}^{} s_{i6}^{} s_{i^{\prime}_{}5}^{} s_{i^{\prime}_{}6}^{}
\cos\left(\beta_{i}^{} + \beta_{i^\prime_{}}^{}\right) \right\}
\nonumber \\
\hspace{-0.25cm} && \hspace{-0.25cm}
+ \hspace{0.06cm} 2 M_4^{} M_5^{2} M_6^{} \left\{ 2 \left[
\sum^3_{i=1} \sum_{i^{\prime} = 1}^3 s_{i4}^{} s_{i5}^{}
s_{i^{\prime}5}^{} s_{i^{\prime}6}^{} \cos\left(\alpha_i^{} +
\beta_{i^\prime_{}}^{}\right)\right]^2 \right.
\nonumber \\
\hspace{-0.25cm} && \hspace{-0.25cm}
\left. - \left[ \sum^3_{i=1} \sum_{i^{\prime} = 1}^3
s_{i4}^{} s_{i5}^{} s_{i^{\prime}_{}4} s_{i^{\prime} 5}
\cos\left(\alpha_i^{} - \alpha_{i^{\prime}_{}}^{}\right)\right] \times
\left[ \sum^3_{i=1} \sum_{i^{\prime} = 1}^3 s_{i5}^{} s_{i6}^{}
s_{i^{\prime}_{}5} s_{i^{\prime} 6} \cos\left(\beta_i^{} -
\beta_{i^{\prime}_{}}^{}\right)\right] \right.
\nonumber \\
\hspace{-0.25cm} && \hspace{-0.25cm}
\left . - \hspace{0.06cm} 2 I_{55}^{}
\sum^3_{i=1} \sum_{i^{\prime} = 1}^3 \sum^3_{i^{\prime\prime}=1}
s_{i4}^{} s_{i5}^{} s_{ i^{\prime}_{} 5}^{} s_{ i^{\prime}_{} 6}^{}
s_{ i^{\prime\prime}_{} 4}^{} s_{i^{\prime\prime}_{} 6}^{}
\cos\left(\alpha_i^{} + \beta_{i_{}^{\prime}} - \gamma_{i_{}^{\prime\prime}}\right)
+ I_{55}^2 \sum^3_{i=1} \sum_{i^{\prime} = 1}^3
s_{i4}^{} s_{i6}^{} s_{i^{\prime}_{}4}^{} s_{i^{\prime}_{}6}^{}
\cos\left(\gamma_{i}^{} + \gamma_{i^\prime_{}}^{}\right) \right\}
\nonumber \\
\hspace{-0.25cm} && \hspace{-0.25cm}
+ \hspace{0.06cm} 2 M_4^{} M_5^{} M_6^{2} \left\{ 2 \left[
\sum^3_{i=1} \sum_{i^{\prime} = 1}^3 s_{i5}^{} s_{i6}^{}
s_{i^{\prime}4}^{} s_{i^{\prime}6}^{} \cos\left(\beta_i^{} +
\gamma_{i^\prime_{}}^{}\right) \right]^2 \right.
\nonumber \\
\hspace{-0.25cm} && \hspace{-0.25cm}
\left. - \left[ \sum^3_{i=1} \sum_{i^{\prime} = 1}^3
s_{i5}^{} s_{i6}^{} s_{i^{\prime}_{}5}
s_{i^{\prime} 6} \cos\left(\beta_i^{} - \beta_{i^{\prime}_{}}^{}\right)\right]
\times \left[ \sum^3_{i=1} \sum_{i^{\prime} = 1}^3 s_{i4}^{} s_{i6}^{}
s_{i^{\prime}_{}4} s_{i^{\prime} 6} \cos\left(\gamma_i^{} -
\gamma_{i^{\prime}_{}}^{}\right)\right] \right.
\nonumber \\
\hspace{-0.25cm} && \hspace{-0.25cm}
\left. - \hspace{0.06cm} 2 I_{66}^{}
\sum^3_{i=1} \sum_{i^{\prime} = 1}^3 \sum^3_{i^{\prime\prime}=1}
s_{i4}^{} s_{i5}^{} s_{ i^{\prime}_{} 5}^{} s_{ i^{\prime}_{} 6}^{}
s_{ i^{\prime\prime}_{} 4}^{} s_{i^{\prime\prime}_{} 6}^{}
\cos\left(\alpha_i^{} - \beta_{i_{}^{\prime}} - \gamma_{i_{}^{\prime\prime}}\right)
+ I_{66}^2 \sum^3_{i=1} \sum_{i^{\prime} = 1}^3
s_{i4}^{} s_{i5}^{} s_{i^{\prime}_{}4}^{} s_{i^{\prime}_{}5}^{}
\cos\left(\alpha_{i}^{} + \alpha_{i^\prime_{}}^{}\right) \right\}
\nonumber \\
\hspace{-0.25cm} && \hspace{-0.25cm}
+ \hspace{0.06cm} M_4^2 M_5^2 \left[ \sum^3_{i=1} \sum_{i^{\prime} = 1}^3
s_{i 4}^{} s_{i 5}^{} s_{i^{\prime}4}^{} s_{i^{\prime} 5}^{}
\cos\left(\alpha_{i}^{} - \alpha_{i^{\prime}}^{}\right) -
I_{44}^{} I_{55}^{} \right]^2
\nonumber \\
\hspace{-0.25cm} && \hspace{-0.25cm}
+ \hspace{0.06cm} M_5^2 M_6^2 \left[ \sum^3_{i=1} \sum_{i^{\prime} = 1}^3
s_{i 5}^{} s_{i 6}^{} s_{i^{\prime}5}^{} s_{i^{\prime} 6}^{}
\cos\left(\beta_{i}^{} - \beta_{i^{\prime}}^{}\right) -
I_{55}^{} I_{66}^{} \right]^2
\nonumber \\
\hspace{-0.25cm} && \hspace{-0.25cm}
+ \hspace{0.06cm} M_4^2 M_6^2 \left[ \sum^3_{i=1} \sum_{i^{\prime} = 1}^3
s_{i 4}^{} s_{i 6}^{} s_{i^{\prime}4}^{} s_{i^{\prime} 6}^{}
\cos\left(\gamma_{i}^{} - \gamma_{i^{\prime}}^{}\right) -
I_{44}^{} I_{66}^{} \right]^2 \; ,
\label{3.11}
\end{eqnarray}
where $I_{jj}^{} = s_{1j}^2 + s_{2j}^2 + s_{3j}^2$ (for $j = 4, 5, 6$)
can be directly read off from Eq.~(\ref{3.3}), and the three sets of
CP-violating phase differences defined as
\begin{eqnarray}
\alpha_i^{} \equiv \delta_{i4}^{} - \delta_{i5}^{} \; , \quad
\beta_i^{} \equiv \delta_{i5}^{} - \delta_{i6}^{} \; , \quad
\gamma_i^{} \equiv \delta_{i6}^{} - \delta_{i4}^{} \;
\label{3.12}
\end{eqnarray}
satisfy the sum rule $\alpha^{}_i + \beta^{}_i + \gamma^{}_i = 0$ (for
$i = 1, 2, 3$). So six of the above nine phase parameters are independent
and serve as the original source of CP violation in the seesaw mechanism.
For instance, one may simply fix $\alpha^{}_i$ and $\beta^{}_i$ (for
$ = 1, 2, 3$) as the six original CP-violating phases.

\subsection{Flavor mixing angles and the effective Dirac phase}

Now let us calculate the three flavor mixing angles
($\theta_{12}^{}$, $\theta_{13}^{}$,
$\theta_{23}^{}$) and the effective Dirac CP-violating phase
($\delta \equiv \delta^{}_{13} - \delta^{}_{12} - \delta^{}_{23}$) of $U^{}_0$
which determine the strengths of flavor oscillations for
the three active neutrinos. A combination of the unitarity of $U^{}_0$ and
 Eq.~(\ref{3.1}) leads us to
\begin{eqnarray}
\left(U_0^{}\right)_{\alpha 1}^{} \left(U_0^{}\right)_{\beta 1}^{*} + 	
\left(U_0^{}\right)_{\alpha 2}^{} \left(U_0^{}\right)_{\beta 2}^{*} + 	
\left(U_0^{}\right)_{\alpha 3}^{} \left(U_0^{}\right)_{\beta 3}^{*}
\hspace{-0.25cm} & = & \hspace{-0.25cm} \delta^{}_{\alpha\beta} \; ,
\nonumber \\
m_1^2 \left(U_0^{}\right)_{\alpha 1}^{} \left(U_0^{}\right)_{\beta 1}^{*} +
m_2^2 \left(U_0^{}\right)_{\alpha 2}^{} \left(U_0^{}\right)_{\beta 2}^{*} +
m_3^2 \left(U_0^{}\right)_{\alpha 3}^{} \left(U_0^{}\right)_{\beta 3}^{*}
\hspace{-0.25cm} & = & \hspace{-0.25cm} H^{}_{\alpha\beta} \; ,
\nonumber \\
m_1^4 \left(U_0^{}\right)_{\alpha 1}^{} \left(U_0^{}\right)_{\beta 1}^{*} +
m_2^4 \left(U_0^{}\right)_{\alpha 2}^{} \left(U_0^{}\right)_{\beta 2}^{*} +
m_3^4 \left(U_0^{}\right)_{\alpha 3}^{} \left(U_0^{}\right)_{\beta 3}^{*}
\hspace{-0.25cm} & = & \hspace{-0.25cm} \left(H^2\right)_{\alpha\beta} \; ,
\hspace{0.8cm}
\label{3.13}
\end{eqnarray}
where $\delta_{\alpha\beta}^{}$ is the Kronecker symbol for
$\alpha,\beta = e, \mu, \tau$. The three variables
$\left(U_0^{}\right)_{\alpha i}^{} \left(U_0^{}\right)_{\beta i}^{*}$
(for $i=1, 2, 3$) can then be derived from the above three linear equations.
In the case of $\alpha = \beta$, we obtain
\begin{eqnarray}
\left|\left(U_0^{}\right)_{\alpha 1}^{}\right|^2
\hspace{-0.25cm} & = & \hspace{-0.25cm}
\frac{\left(H^2\right)_{\alpha\alpha}^{} - H_{\alpha\alpha}^{} \left(m_2^2
+ m_3^2\right) + m_2^2 m_3^2}{\Delta_{21}^{} \Delta_{31}} \; ,
\nonumber \\
\left|\left(U_0^{}\right)_{\alpha 2}^{}\right|^2
\hspace{-0.25cm} & = & \hspace{-0.25cm}
\frac{\left(H^2\right)_{\alpha\alpha}^{} - H_{\alpha\alpha}^{} \left(m_1^2
+ m_3^2\right) + m_1^2 m_3^2}{\Delta_{21}^{} \Delta_{23}} \; ,
\nonumber \\
\left|\left(U_0^{}\right)_{\alpha 3}^{}\right|^2
\hspace{-0.25cm} & = & \hspace{-0.25cm}
\frac{\left(H^2\right)_{\alpha\alpha}^{} - H_{\alpha\alpha}^{} \left(m_1^2
+ m_2^2\right) + m_1^2 m_2^2}{\Delta_{31}^{}\Delta_{32}} \; ; \hspace{0.8cm}
\label{3.14}
\end{eqnarray}
and in the case of $\alpha \neq \beta$, we arrive at
\begin{eqnarray}
\left(U_0^{}\right)_{\alpha 1}^{} \left(U_0^{}\right)_{\beta 1}^{*}
\hspace{-0.25cm} & = & \hspace{-0.25cm}
\frac{\left(H^2\right)_{\alpha\beta} - H_{\alpha\beta}^{} \left(m_2^2 +
m_3^2\right)}{\Delta_{21}\Delta_{31}} \; ,
\nonumber \\
\left(U_0^{}\right)_{\alpha 2}^{} \left(U_0^{}\right)_{\beta 2}^{*}
\hspace{-0.25cm} & = & \hspace{-0.25cm}
\frac{\left(H^2\right)_{\alpha\beta} - H_{\alpha\beta}^{} \left(m_1^2 +
m_3^2\right)}{\Delta_{21}\Delta_{23}} \; ,
\nonumber\\
\left(U_0^{}\right)_{\alpha 3}^{} \left(U_0^{}\right)_{\beta 3}^{*}
\hspace{-0.25cm} & = & \hspace{-0.25cm}
\frac{\left(H^2\right)_{\alpha\beta} - H_{\alpha\beta}^{} \left(m_1^2 +
m_2^2\right)}{\Delta_{31}\Delta_{32}} \; . \hspace{0.9cm}
\label{3.15}
\end{eqnarray}
In view of Eq.~(\ref{1.2}), one finds that the three flavor mixing angles are
related to the moduli of $U^{}_0$ in its first row and third column as follows:
\begin{eqnarray}
s_{12}^2 = \frac{\left|\left(U_0^{}\right)_{e2}\right|^2}{1 -
\left|\left(U_0^{}\right)_{e3}\right|^2} \; , \quad	
s_{13}^2 = \left|\left(U_0^{}\right)_{e3}\right|^2 \; , \quad
s_{23}^2 = \frac{\left|\left(U_0^{}\right)_{\mu 3}\right|^2}{1 -
\left|\left(U_0^{}\right)_{e3}\right|^2} \; .
\label{3.16}
\end{eqnarray}
Substituting Eq.~(\ref{3.14}) into Eq.~(\ref{3.16}), we get
\begin{eqnarray}
s_{12}^2 \hspace{-0.25cm} & = & \hspace{-0.25cm}
\frac{\left[\left(H^2\right)_{ee}^{} - H_{ee}^{} \left(m_1^2 + m_3^2\right)
+ m_1^2 m_3^2 \right] \Delta_{31}^{}}
{\left[\left(H^2\right)_{ee}^{} - H_{ee}^{} \left(m_1^2 + m_2^2\right)
+ m_1^2 m_2^2 -\Delta_{31}^{} \Delta_{32}^{} \right] \Delta_{21}^{}} \; , \hspace{0.6cm}
\nonumber \\
s_{13}^2 \hspace{-0.25cm} & = & \hspace{-0.25cm}
\frac{\left(H^2\right)_{ee}^{} - H_{ee}^{} \left(m_1^2 + m_2^2\right) +
m_1^2 m_2^2}{\Delta_{31}^{}\Delta_{32}^{}} \; ,
\nonumber \\
s_{23}^2 \hspace{-0.25cm} & = & \hspace{-0.25cm}
\frac{\left(H^2\right)_{\mu\mu}^{} - H_{\mu\mu}^{} \left(m_1^2 + m_2^2\right)
+ m_1^2 m_2^2}{- \left(H^2\right)_{ee}^{} + H_{ee}^{} \left(m_1^2 + m_2^2\right)
- m_1^2 m_2^2 + \Delta_{31}^{}\Delta_{32}^{}} \; ,
\label{3.17}
\end{eqnarray}
where the explicit expressions of $H_{\alpha\alpha}^{}$ and
$\left(H^2\right)_{\alpha\alpha}^{}$ can be found in Appendix A, and
those of $m_i^2$ and $\Delta_{ii^{\prime}_{}}$ (for $i, i^\prime = 1, 2, 3$) can be
achieved from Eqs. (\ref{3.5})---(\ref{3.11}).

On the other hand, the effective Dirac CP-violating phase $\delta$ can be extracted
from the Jarlskog invariant ${\cal J}^{}_\nu$ shown in Eq.~(\ref{1.5}).
Namely
\footnote{Of course, one may also determine $\cos\delta$ from
$|\left(U^{}_0\right)^{}_{\mu 1}|^2$,
$|\left(U^{}_0\right)^{}_{\mu 2}|^2$,
$|\left(U^{}_0\right)^{}_{\tau 1}|^2$ or
$|\left(U^{}_0\right)^{}_{\tau 2}|^2$ by use of the cosine theorem and with the help
of Eqs.~(\ref{3.14}) and (\ref{3.17}).},
\begin{eqnarray}
\sin\delta = \frac{{\cal J}_{\nu}^{}}{\sqrt{s_{12}^2 \hspace{0.05cm}
s_{13}^2 \hspace{0.05cm} s_{23}^2
\left(1- s_{12}^2\right) \left(1- s_{23}^2\right)} \hspace{0.05cm}
\left(1-s_{13}^2\right)} \; ,
\label{3.18}
\end{eqnarray}
where the expressions of $s^2_{i i^\prime}$ (for $i i^\prime = 12, 13, 23$)
have been given in Eq.~(\ref{3.17}), and ${\cal J}^{}_\nu$ has already been
calculated in Ref.~\cite{Xing:2024xwb} with the help of
\begin{eqnarray}
{\cal J_{\nu}^{}} = \frac{\displaystyle {\rm Im}\left[\sum_{\gamma}
\epsilon_{\alpha\beta\gamma}^{} H_{\alpha \beta}^{}
\left(H^2\right)_{\beta\alpha}\right]}
{\Delta_{21}^{}\Delta_{31}^{}\Delta_{32}^{}}
= \frac{{\rm Im}\left(H_{e\mu}^{} H_{\mu \tau}^{} H_{\tau e}^{}\right)}{\Delta_{21}^{}\Delta_{31}^{}\Delta_{32}^{}} \;
\label{3.19}
\end{eqnarray}
for $(\alpha, \beta, \gamma)$ running cyclically over $(e, \mu, \tau)$. The
explicit analytical result of ${\cal J}_{\nu}^{}$ consists of the terms
proportional to $M_j^{3} M_k^3$, $M_j^{4} M_k^2$, $M_j^{4} M_k^{} M_l^{}$,
$M_j^{3} M_k^{2} M_l^{}$ and $M_j^{2} M_k^{2} M_l^{2}$ (for
$j \neq k \neq l = 4, 5, 6$)~\cite{Xing:2024xwb}, and the product
$\Delta_{21}^{}\Delta_{31}^{}\Delta_{32}^{}$ can be
directly deduced from Eq.~(\ref{3.8}) or Eq.~(\ref{3.9}).

Note that ${\cal J}^{}_\nu$ depends on the six original and independent
phase parameters of the seesaw mechanism (e.g., $\alpha^{}_i$ and $\beta^{}_i$
for $i = 1, 2, 3$), as specifically shown in Ref.~\cite{Xing:2024xwb}. So
$\sin\delta$ must also be a function of $\sin\alpha^{}_i$ and $\sin\beta^{}_i$,
similar to ${\cal J}^{}_\nu$. The point is that all the CP-violating phases
in the seesaw mechanism are of the Majorana nature, in the sense that they
may manifest themselves in those lepton-number-violating processes. That is
why $\delta$ is referred to as the ``effective" Dirac CP-violating phase
which is unique to CP violation in neutrino oscillations, given the fact that
non-unitarity of the PMNS matrix $U$ is strongly suppressed and realistically
negligible in the foreseeable neutrino experiments.

It is the {\it first} time that the general and explicit analytical
relations between the 6 {\it derivational} neutrino oscillation parameters and
the 18 {\it original} seesaw flavor parameters have been established.
The approximation made in our calculations is equivalent to concentrating on
the unique Weinberg dimension-5 operator~\cite{Weinberg:1979sa} in the
seesaw-extended SMEFT, simply because small
non-unitarity of $U$ is associated with those dimension-6 operators of the SMEFT
and its effects are almost experimentally untouchable in the near future. Our
results are expected to be useful for testing the seesaw mechanism in a variety of
existing and upcoming neutrino oscillation experiments.

Here let us make some brief comments on a few well-known non-oscillation
quantities of massive neutrinos in the canonical seesaw framework.
\begin{itemize}
\item     It is straightforward to calculate the sum of three light
neutrino masses $\Sigma^{}_\nu \equiv m^{}_1 + m^{}_2 + m^{}_3$ with
the help of Eq.~(\ref{3.5}) or Eq.~(\ref{3.6}), and its upper bound extracted
from a number of cosmological observations~\cite{Planck:2018vyg} can therefore
be used to constrain the original seesaw parameter space.

\item     The effective mass of the beta decays, namely
$\langle m\rangle^{}_e \equiv \sqrt{m_1^{2} \left|U_{e1}^{}\right|^2 + m_2^{2}
\left|U_{e2}^{}\right|^2 + m_3^{2} \left|U_{e3}^{}\right|^2}$, can also be used
to constrain the parameter space of the seesaw mechanism, since its upper bound
has recently been obtained from the KATRIN experiment~\cite{Fengler:2024ufb}.
Note that $\langle m\rangle^{}_e = \sqrt{H^{}_{ee}}$ holds, as can be seen from
Eq.~(\ref{3.13}). The explicit expression of $H^{}_{ee}$ is given
by Eq.~(\ref{A.1}) in Appendix A.

\item     The effective mass of the neutrinoless double-beta decays,
denoted as $\langle m\rangle^{}_{0\nu2\beta}$, is composed of
both the $m^{}_i U^2_{e i}$ terms (for $i = 1, 2, 3$)
and the $R^2_{e j}/M^{}_j$ terms (for $j = 4, 5, 6$) in the
canonical seesaw framework~\cite{Xing:2020ijf}, as such
lepton-number-violating processes are mediated by both light and heavy
Majorana neutrinos. Given
$\left(m^{}_1 U^2_{e 1} + m^{}_2 U^2_{e 2} + m^{}_3 U^2_{e 3}\right)
= -\left(M^{}_4 R^2_{e 4} + M^{}_5 R^2_{e 5} + M^{}_6 R^2_{e 6}\right)$ as
indicated by the exact seesaw relation in Eq.~(\ref{1.3}),
we find that $\langle m\rangle^{}_{0\nu2\beta}$
depends only on the original seesaw flavor parameters $M^{}_j$ and $R^{}_{e j}$.
So the upper bound of $\langle m\rangle^{}_{0\nu2\beta}$~\cite{Agostini:2022zub}
is certainly useful to constrain the seesaw parameter space.
\end{itemize}
It is also helpful to consider those charged-lepton-flavor-violating (cLFV) 
processes, such as the radiative decays of $\mu$ and $\tau$ as well as the
$\mu \to e$ conversions, to probe or constrain the parameter space of the 
seesaw mechanism (e.g., as already done in 
Refs.~\cite{Blennow:2023mqx,Xing:2020ivm}).
It is remarkable that such cLFV processes may provide the direct and 
stringent constraints on non-unitarity of the PMNS matrix $U$; i.e., 
the deviation of $U=A U_0^{} = \left(I-\alpha \right)U_0^{}$ from
$U^{}_0$, where $A$ is defined in Eq. \eqref{2.5} and consists of all
the original active-sterile flavor mixing parameters. The
cLFV processes can therefore be used to directly constrain the off-diagonal 
elements $\alpha_{ii^\prime}^{}$ (for $i\neq i^\prime =1,2,3$), and the
relevant results can further be translated to the constraints on the 
diagonal elements $\alpha_{ii}^{}$ (for $i=1,2,3$) by means of the 
Schwartz inequality \cite{Blennow:2023mqx}. Of course, a careful global
analysis of all the available experimental information on the PMNS matrix 
$U$ at low energies will definitely be needed, in order to probe the original 
seesaw parameter space as much as possible in the upcoming precision 
measurement era.

\setcounter{equation}{0}
\section{Results for the minimal seesaw scenario}

The minimal seesaw scenario is a simplified version of the canonical seesaw
mechanism which contains only two right-handed neutrino 
fields~\cite{Kleppe:1995zz,
Ma:1998zg,Frampton:2002qc,Endoh:2002wm, 
Raidal:2002xf,Dutta:2003ps,Ibarra:2003xp,Xing:2020ald} and thus involves 
only 11 original
flavor parameters (i.e., 2 heavy neutrino masses, 6 active-sterile flavor mixing
angles and 3 independent CP-violating phases~\cite{Xing:2023kdj}). The most salient
feature of this interesting and instructive scenario is $m^{}_1 = 0$ in the
NMO case or $m^{}_3 = 0$ in the IMO case, as a straightforward consequence of switching
off the third heavy Majorana neutrino field $N_6^{}$ and its relevant flavor parameters.
The relevant results obtained in section 3 can therefore be simplified to a great extent.

\begin{figure}[t]
\begin{center}
\includegraphics[width=13.35cm]{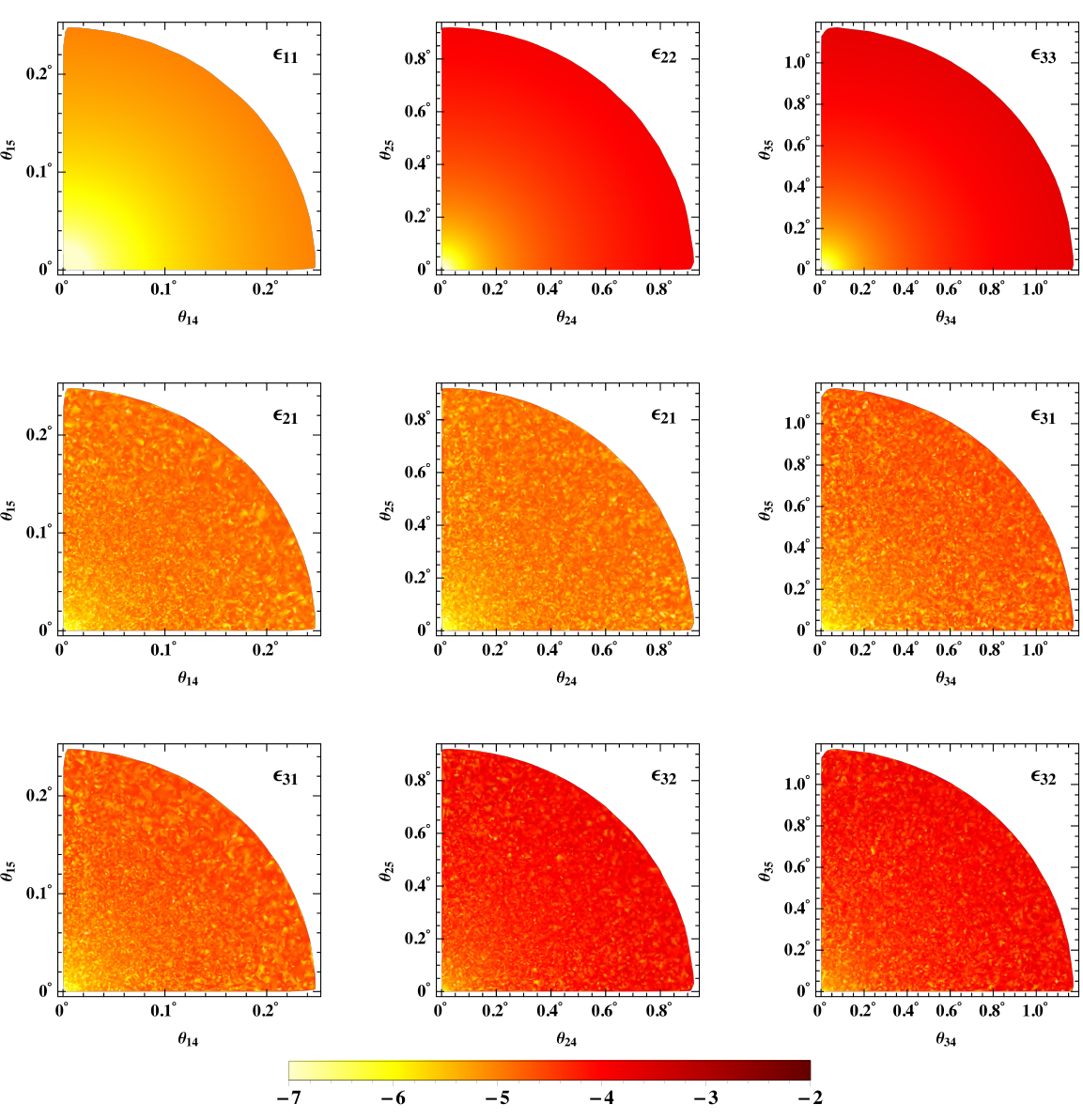}
\caption{Numerical constraints on the active-sterile flavor mixing angles
$\theta^{}_{1j}$, $\theta^{}_{2j}$ and $\theta^{}_{3j}$ (for $j = 4, 5$)
with the help of the upper bounds on $\epsilon^{}_{ii^\prime}$ (for
$i, i^\prime = 1, 2, 3$) in Eq.~(\ref{4.1}) in the NMO case of the minimal
seesaw scenario, where different colors in the bar legend denote different
values of $\log_{10}\left(\epsilon^{}_{i i^{\prime}}\right)$.}
\label{fig3}
\end{center}
\end{figure}

\begin{figure}[t]
\begin{center}
\includegraphics[width=13.35cm]{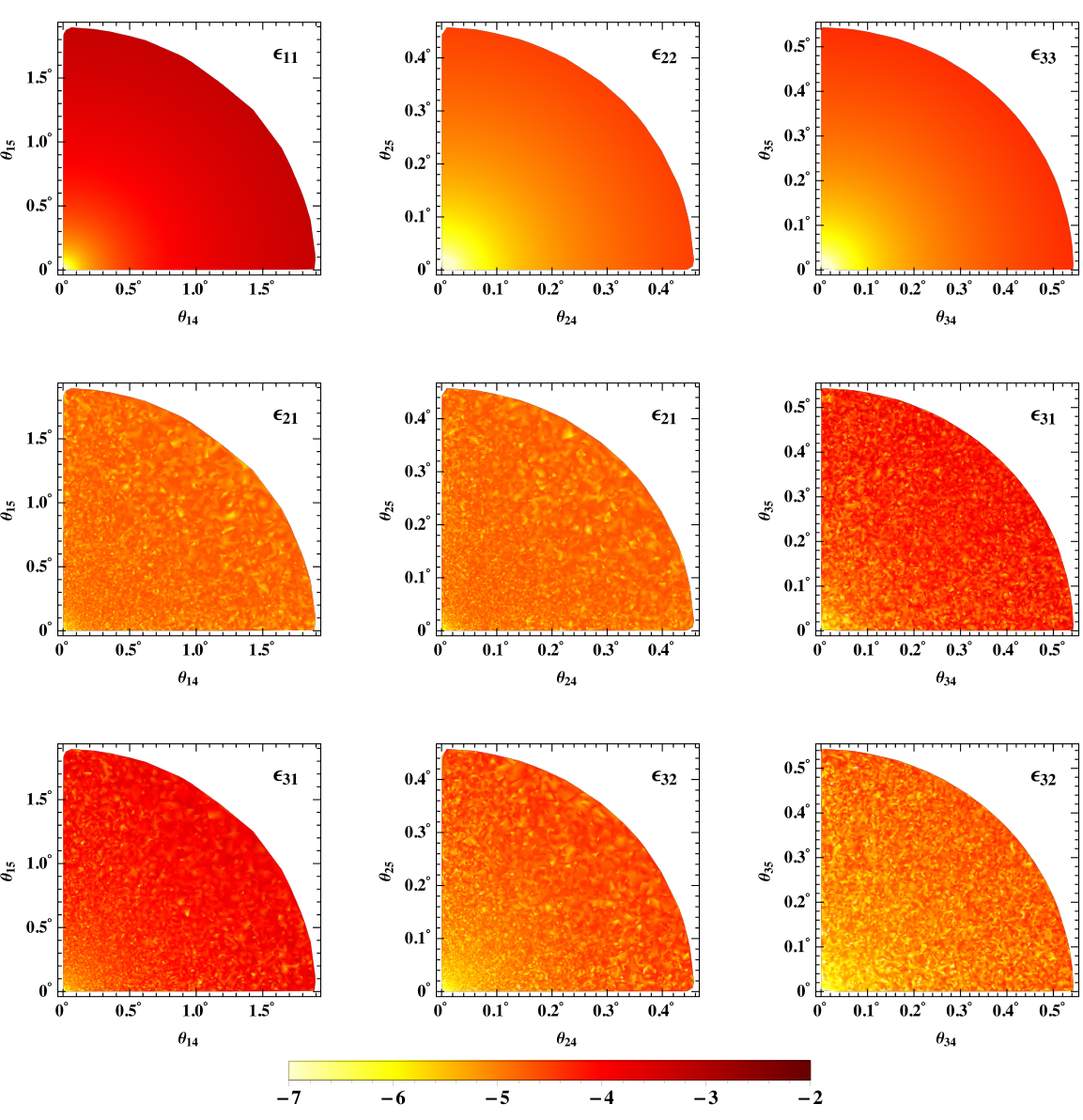}
\caption{Numerical constraints on the active-sterile flavor mixing angles
$\theta^{}_{1j}$, $\theta^{}_{2j}$ and $\theta^{}_{3j}$ (for $j = 4, 5$)
with the help of the upper bounds on $\epsilon^{}_{ii^\prime}$ (for
$i, i^\prime = 1, 2, 3$) in Eq.~(\ref{4.1}) in the IMO case of the minimal
seesaw scenario, where different colors in the bar legend denote different
values of $\log_{10}\left(\epsilon^{}_{i i^{\prime}}\right)$.}
\label{fig4}
\end{center}
\end{figure}

Note that the experimental constraints on the active-sterile flavor mixing angles
in the minimal seesaw framework are much stronger than in the canonical case. Taking
account of the latest global-fit results about non-unitarity of the PMNS matrix
$U$ obtained in Ref.~\cite{Blennow:2023mqx}, namely
\begin{eqnarray}
\epsilon \equiv \left| I - A\right| < \left\{
\begin{array}{l}
\left( \begin{matrix}
9.4 \times 10^{-6} & 0 & 0 \cr
2.4 \times 10^{-5} & 1.3 \times 10^{-4} & 0 \cr
4.4 \times 10^{-5} & 2.6 \times 10^{-4} & 2.1 \times 10^{-4} \cr \end{matrix}
\right) \;  \hspace{0.6cm} ({\rm NMO}) \;
\\ \vspace{-0.45cm} \\
\left( \begin{matrix}
5.5 \times 10^{-4} & 0 & 0 \cr
2.6 \times 10^{-5} & 3.2 \times 10^{-5} & 0 \cr
2.8 \times 10^{-4} & 7.0 \times 10^{-5} & 4.5 \times 10^{-5} \cr \end{matrix}
\right) \;  \hspace{0.6cm} ({\rm IMO}) \;
\end{array} \right.
\label{4.1}
\end{eqnarray}
in the minimal seesaw scenario, we switch off
$\theta^{}_{i6}$ and $\delta^{}_{i6}$ in Eq.~(\ref{2.1})
and confront the resulting $A$ with Eq.~(\ref{4.1}) for arbitrary values of
$\delta^{}_{i4}$ and $\delta^{}_{i5}$ (for $i = 1, 2, 3$)
\footnote{Note that $R$ is actually a $3\times 2$ matrix in the minimal seesaw
framework, and thus the rank of $D^{}_\nu$ in Eq.~(\ref{1.3}) must be two,
implying that one of the three light Majorana neutrinos has to be massless at the tree
level.}.
Figs.~\ref{fig3} and \ref{fig4} illustrate how the six nonzero elements
of $\epsilon$ in Eq.~(\ref{4.1}) depend on $\theta_{i4}^{}$ and $\theta_{i5}^{}$
(for $i=1,2,3$) in the respective NMO and IMO cases, and the upper bounds of
these six flavor mixing angles are extracted as follows:
\begin{eqnarray}
&& \theta^{}_{1j} < 0.25^\circ \; , \quad
\theta^{}_{2j} < 0.92^\circ \; , \quad
\theta^{}_{3j} < 1.17^\circ \;
\hspace{0.8cm} ({\rm NMO}) \; ; \hspace{1cm}
\nonumber \\
&& \theta^{}_{1j} < 1.90^\circ \; , \quad
\theta^{}_{2j} < 0.46^\circ \; , \quad
\theta^{}_{3j} < 0.54^\circ \;
\hspace{0.8cm} ({\rm IMO}) \; ,
\label{4.2}
\end{eqnarray}
where $j = 4$ and $5$. These stringent constraints strongly support our
analytical approximations made in this work --- only keeping the leading
terms of $A$ and $R$ in our calculations.

In the minimal seesaw scenario, the analytical expressions of neutrino
masses can be simplified to
\begin{eqnarray}
m^2_{2,3} \hspace{-0.25cm} & = & \hspace{-0.25cm}
\frac 12 \left(x^\prime \mp \sqrt{y^\prime}\right) \; ,
\nonumber \\
\Delta_{21}^{} \hspace{-0.25cm} & = & \hspace{-0.25cm}
\frac 12 \left(x^\prime - \sqrt{y^\prime}\right) \; , \quad
\Delta_{31}^{} =
\frac 12 \left(x^\prime + \sqrt{y^\prime}\right) \; , \quad
\Delta_{32}^{} = \sqrt{y^\prime} \hspace{0.6cm}
\label{4.3}
\end{eqnarray}
in the NMO case; or
\begin{eqnarray}
m^2_{1,2} \hspace{-0.25cm} & = & \hspace{-0.25cm}
\frac 12 \left(x^\prime \mp \sqrt{y^\prime}\right) \; ,
\nonumber \\
\Delta_{21}^{} \hspace{-0.25cm} & = & \hspace{-0.25cm}
\sqrt{y^\prime} \; , \quad
\Delta_{13}^{} =
\frac 12 \left(x^\prime - \sqrt{y^\prime}\right) \; , \quad
\Delta_{23}^{} =
\frac 12 \left(x^\prime + \sqrt{y^\prime}\right) \; \hspace{0.6cm}
\label{4.4}
\end{eqnarray}
in the IMO case, where $x^\prime \equiv {\rm tr}\left(H\right)$ and
$y^\prime \equiv 2  {\rm tr}\left(H^2\right) - \left[{\rm tr}
\left(H\right)\right]^2$. The expressions of $x^\prime$ and $y^\prime$ are
\begin{eqnarray}
x^\prime \hspace{-0.25cm} & = & \hspace{-0.25cm}
M_4^2 I_{44}^2 + M_5^2 I_{55}^2 + 2 M_4^{} M_5^{}
\sum^3_{i=1} \sum^3_{i^{\prime}=1}
s_{i4}^{} s_{i5}^{} s_{i^{\prime}_{}4}^{} s_{i^{\prime}_{}5}^{}
\cos\left(\alpha_i^{} + \alpha_{i^{\prime}_{}}^{}\right) \; ,
\nonumber \\
y^\prime \hspace{-0.25cm} & = & \hspace{-0.25cm}
M_4^4  I_{44}^4+ M_5^4 I_{55}^4 + 4 M_4^{} M_5^{}
\left(M_4^2 I_{44}^2  + M_5^2 I_{55}^2 \right) \times
\left[\sum^3_{i=1} \sum^3_{i^{\prime}=1}
s_{i4}^{} s_{i5}^{}  s_{i^{\prime}_{}4}^{} s_{i^{\prime}_{}5}^{}
\cos\left(\alpha_i^{} + \alpha_{i^{\prime}_{}}^{}\right) \right] \hspace{0.6cm}
\nonumber \\
\hspace{-0.25cm} && \hspace{-0.25cm}	
- \hspace{0.06cm} 2 M_4^2 M_5^2 \Bigg\{ I_{44}^2 I_{55}^2 -
4 I_{44}^{} I_{55}^{} \sum^3_{i=1} \sum^3_{i^{\prime}=1} s_{i4}^{} s_{i5}^{} s_{i^{\prime}_{}4}^{} s_{i^{\prime}_{}5}^{}
\cos\left(\alpha_i^{} - \alpha_{i^{\prime}_{}}^{}\right)
\nonumber \\
\hspace{-0.25cm} && \hspace{-0.25cm}	
+ \hspace{0.06cm} 2 \left[\sum^3_{i=1} \sum^3_{i^{\prime}=1}
s_{i4}^{} s_{i5}^{}  s_{i^{\prime}_{}4}^{} s_{i^{\prime}_{}5}^{}
\sin\left(\alpha_i^{} + \alpha_{i^{\prime}_{}}^{}\right)\right]^2
\Bigg\} \; ,
\label{4.5}
\end{eqnarray}
in which $I^{}_{jj} = s^2_{1j} + s^2_{2j} + s^2_{3j}$ (for $j = 4, 5$). So
switching off the 6 active-sterile flavor mixing angles in the minimal
seesaw framework, which is equivalent to switching off the Yukawa interactions
between left- and right-handed neutrinos, will immediately lead us to
$x^\prime =0$, $y^\prime =0$ and thus $m^{}_1 = m^{}_2 = m^{}_3 = 0$.

The general results for the three active flavor mixing angles obtained in
Eq.~(\ref{3.17}) can also be simplified by taking $m^{}_1 = 0$ in the NMO case
or $m^{}_3 = 0$ in the IMO case. In either case the relevant expressions of
$H_{\alpha\alpha}^{}$ and $\left(H^2\right)_{\alpha\alpha}^{}$ (for
$\alpha = e, \mu, \tau$) listed in Appendix A are simplified to
\begin{eqnarray}
H_{ee}^{} \hspace{-0.25cm} & = & \hspace{-0.25cm}
M_4^2 I_{44}^{} s_{14}^2 + M_5^2 I_{55}^{} s_{15}^2 + 2 M_4^{} M_5^{}
s_{14}^{} s_{15}^{} \left[\sum_{i=1}^3 s_{i4}^{} s_{i5}^{}
\cos\left(\alpha_i^{} + \alpha_1^{}\right)\right] \; ,
\nonumber \\
H_{\mu\mu}^{} \hspace{-0.25cm} & = & \hspace{-0.25cm}
M_4^2 I_{44}^{} s_{24}^2 + M_5^2 I_{55}^{} s_{25}^2 + 2 M_4^{}
M_5^{} s_{24}^{} s_{25}^{} \left[\sum_{i=1}^3 s_{i4}^{} s_{i5}^{}
\cos\left(\alpha_i^{} + \alpha_2^{}\right)\right] \; ,
\nonumber \\
H_{\tau\tau}^{} \hspace{-0.25cm} & = & \hspace{-0.25cm}
M_4^2 I_{44}^{} s_{34}^2 + M_5^2 I_{55}^{} s_{35}^2 + 2 M_4^{}
M_5^{} s_{34}^{} s_{35}^{} \left[\sum_{i=1}^3 s_{i4}^{} s_{i5}^{}
\cos\left(\alpha_i^{} + \alpha_3^{}\right)\right] \; , \hspace{0.6cm}
\label{4.6}
\end{eqnarray}
and
\begin{eqnarray}
\left(H^2\right)_{ee}^{} \hspace{-0.25cm} & = & \hspace{-0.25cm}
M_4^4 I_{44}^3 s_{14}^2 + M_5^4 I_{55}^3 s_{15}^2 + 2 M_4^{}
M_5^{} \Bigg\{ s_{14}^{} s_{15}^{} \left(M_4^2 I_{44}^2 + M_5^2 I_{55}^2\right)
\left[\sum_{i=1}^3 s_{i4}^{} s_{i5}^{}
\cos\left(\alpha_i^{} + \alpha_1^{}\right) \right] \hspace{0.6cm}
\nonumber \\
\hspace{-0.25cm} && \hspace{-0.25cm}
+ \hspace{0.06cm} \left( M_4^2 I_{44}^{} s_{14}^2 + M_5^2 I_{55}^{} s_{15}^2\right)
\left[\sum^3_{i=1} \sum^3_{i^{\prime} =1}
s_{i4}^{} s_{i5}^{} s_{i^{\prime}_{}4}^{} s_{i^{\prime}_{}5}^{}
\cos\left(\alpha_i^{} + \alpha_{i^{\prime}_{}}^{}\right)\right]\Bigg\}
\nonumber \\
\hspace{-0.25cm} && \hspace{-0.25cm}
+ \hspace{0.06cm} M_4^2 M_5^2  \Bigg\{ 2 s_{14}^{} s_{15}^{}
\left[\sum^3_{i=1} \sum^3_{i^{\prime} =1} \sum^3_{i^{\prime\prime} =1}
s_{i4}^{} s_{i5}^{} s_{ i^{\prime}_{}4}^{} s_{ i^{\prime}_{}5}^{}
s_{ i^{\prime\prime}_{}4}^{} s_{ i^{\prime\prime}_{}5}^{}
\cos\left(\alpha_i^{} + \alpha_{i^{\prime}_{}}^{} +
\alpha_{i^{\prime\prime}_{}}^{} + \alpha_1^{}\right) \right]
\nonumber \\
\hspace{-0.25cm} && \hspace{-0.25cm}
+ \hspace{0.06cm} \left(I_{44}^{} s_{15}^2 + I_{55}^{} s_{14}^2\right)
\left[\sum^3_{i=1} \sum^3_{i^{\prime}_{} =1}
s_{i4}^{} s_{i5}^{} s_{i^{\prime}_{}4}^{} s_{i^{\prime}_{}5}^{}
\cos\left(\alpha_i^{} - \alpha_{i^{\prime}_{}}^{}\right)\right]
\nonumber \\
\hspace{-0.25cm} && \hspace{-0.25cm}
+ \hspace{0.06cm} 2 s_{14}^{} s_{15}^{} I_{44}^{} I_{55}^{}
\left[\sum_{i=1}^3 s_{i4}^{} s_{i5}^{} \cos\left(\alpha_i^{} -\alpha_1^{}\right)\right]
\Bigg\} \; ,
\nonumber\\
\left(H^2\right)_{\mu\mu}^{} \hspace{-0.25cm} & = & \hspace{-0.25cm}
M_4^4 I_{44}^3 s_{24}^2 + M_5^4 I_{55}^3 s_{25}^2
+ 2 M_4^{} M_5^{} \Bigg\{ s_{24}^{} s_{25}^{}
\left(M_4^2 I_{44}^2 + M_5^2 I_{55}^2\right)
\left[\sum_{i=1}^3 s_{i4}^{} s_{i5}^{} \cos\left(\alpha_i^{} + \alpha_2^{}
\right)\right]
\nonumber \\
\hspace{-0.25cm} && \hspace{-0.25cm}
+ \hspace{0.06cm} \left( M_4^2 I_{44}^{} s_{24}^2 + M_5^2 I_{55}^{} s_{25}^2\right)
\left[\sum^3_{i=1} \sum^3_{i^{\prime} =1} s_{i4}^{} s_{i5}^{} s_{i^{\prime}_{}4}^{}
s_{i^{\prime}_{}5}^{} \cos\left(\alpha_i^{} + \alpha_{i^{\prime}_{}}^{}\right)
\right]\Bigg\}
\nonumber \\
\hspace{-0.25cm} && \hspace{-0.25cm}
+ \hspace{0.06cm} M_4^2 M_5^2 \Bigg\{ 2 s_{24}^{} s_{25}^{}
\left[\sum^3_{i=1} \sum^3_{i^{\prime} =1} \sum^3_{i^{\prime\prime} =1}
s_{i4}^{} s_{i5}^{} s_{ i^{\prime}_{}4}^{} s_{ i^{\prime}_{}5}^{}
s_{ i^{\prime\prime}_{}4}^{} s_{ i^{\prime\prime}_{}5}^{}
\cos\left(\alpha_i^{} + \alpha_{i^{\prime}_{}}^{} + \alpha_{i^{\prime\prime}_{}}^{}
+ \alpha_2^{}\right) \right]
\nonumber \\
\hspace{-0.25cm} && \hspace{-0.25cm}
+ \hspace{0.06cm} \left(I_{44}^{} s_{25}^2+ I_{55}^{} s_{24}^2\right)
\left[\sum^3_{i=1} \sum^3_{i^{\prime} =1} s_{i4}^{} s_{i5}^{} s_{i^{\prime}_{}4}^{} s_{i^{\prime}_{}5}^{} \cos\left(\alpha_i^{} - \alpha_{i^{\prime}_{}}^{}\right)
\right]
\nonumber \\
\hspace{-0.25cm} && \hspace{-0.25cm}
+ \hspace{0.06cm} 2 s_{24}^{} s_{25}^{} I_{44}^{} I_{55}^{}
\left[\sum_{i=1}^3 s_{i4}^{} s_{i5}^{} \cos\left(\alpha_i^{} -\alpha_2^{}\right)
\right] \Bigg\} \; ,
\nonumber \\
\left(H^2\right)_{\tau\tau}^{} \hspace{-0.25cm} & = & \hspace{-0.25cm}
M_4^4 I_{44}^3 s_{34}^2 + M_5^4 I_{55}^3 s_{35}^2
+ 2 M_4^{} M_5^{} \Bigg\{ s_{34}^{} s_{35}^{}
\left(M_4^2 I_{44}^2 + M_5^2 I_{55}^2\right)
\left[\sum_{i=1}^3 s_{i4}^{} s_{i5}^{} \cos\left(\alpha_i^{} + \alpha_3^{}\right)
\right]
\nonumber \\
\hspace{-0.25cm} && \hspace{-0.25cm}
+ \hspace{0.06cm} \left( M_4^2 I_{44}^{} s_{34}^2 + M_5^2 I_{55}^{} s_{35}^2\right)
\left[\sum^3_{i=1} \sum^3_{i^{\prime} =1} s_{i4}^{} s_{i5}^{}
s_{i^{\prime}_{}4}^{} s_{i^{\prime}_{}5}^{} \cos\left(\alpha_i^{} +
\alpha_{i^{\prime}_{}}^{}\right)\right] \Bigg\}
\nonumber \\
\hspace{-0.25cm} && \hspace{-0.25cm}
+ \hspace{0.06cm} M_4^2 M_5^2 \Bigg\{ 2 s_{34}^{} s_{35}^{}
\left[\sum^3_{i=1} \sum^3_{i^{\prime} =1} \sum^3_{i^{\prime\prime} =1}
s_{i4}^{} s_{i5}^{} s_{ i^{\prime}_{}4}^{} s_{ i^{\prime}_{}5}^{}
s_{ i^{\prime\prime}_{}4}^{} s_{ i^{\prime\prime}_{}5}^{} \cos\left(\alpha_i^{}
+ \alpha_{i^{\prime}_{}}^{} + \alpha_{i^{\prime\prime}_{}}^{} + \alpha_3^{}\right)
\right]
\nonumber \\
\hspace{-0.25cm} && \hspace{-0.25cm}
+ \hspace{0.06cm} \left(I_{44}^{} s_{35}^2+ I_{55}^{} s_{34}^2\right)
\left[\sum^3_{i=1} \sum^3_{i^{\prime} =1} s_{i4}^{} s_{i5}^{}
s_{i^{\prime}_{}4}^{} s_{i^{\prime}_{}5}^{} \cos\left(\alpha_i^{} -
\alpha_{i^{\prime}_{}}^{}\right)\right]
\nonumber \\
\hspace{-0.25cm} && \hspace{-0.25cm}
+ \hspace{0.06cm} 2 s_{34}^{} s_{35}^{} I_{44}^{} I_{55}^{}
\left[\sum_{i=1}^3 s_{i4}^{} s_{i5}^{} \cos\left(\alpha_i^{} -\alpha_3^{}\right)
\right] \Bigg\} \; .
\label{4.7}
\end{eqnarray}
It is obvious that $H_{\mu\mu}^{}$ and $\left(H^2\right)_{\mu\mu}^{}$
can respectively be read off from $H_{ee}^{}$ and $\left(H^2\right)_{ee}^{}$
by doing the replacements $\theta_{14}^{} \leftrightarrow \theta_{24}^{}$,
$\theta_{15}^{} \leftrightarrow \theta_{25}^{}$ and $\alpha_1^{} \leftrightarrow \alpha_2^{}$.
Similarly, one may write out $H_{\tau\tau}^{}$ and
$\left(H^2\right)_{\tau\tau}^{}$ from $H_{ee}^{}$ and
$\left(H^2\right)_{ee}^{}$ with the replacements
$\theta_{14}^{} \leftrightarrow \theta_{34}^{}$, $\theta_{15}^{} \leftrightarrow \theta_{35}^{}$ and
$\alpha_1^{} \leftrightarrow \alpha_3^{}$.

Furthermore, the effective Dirac CP-violating phase can be deduced from
Eq.~(\ref{3.18}), where the Jarlskog invariant ${\cal J}_{\nu}^{}$ in
the minimal seesaw scenario is of a much simpler form
\begin{eqnarray}
{\cal J}_{\nu}^{} \hspace{-0.25cm} & = & \hspace{-0.25cm}
\frac{M_4^2 M_5^2 \left(I_{44}^{} I_{55}^{} - |I_{45}^{}|^2\right)}
{\Delta_{21}^{} \Delta_{31}^{} \Delta_{32}^{}}
\Biggl\{ \sum_{i=1}^3 s_{i4}^{} s_{i5}^{} s_{i^{\prime} 4}^{} s^{}_{i^{\prime}5}
\left(M_4^2 I_{44}^{} s_{i^{\prime\prime}4}^2 - M_5^2 I^{}_{55}
s^2_{i^{\prime\prime} 5}\right) \sin\left(\alpha_i^{} -
\alpha_{i^{\prime}}^{}\right) \hspace{0.6cm}
\nonumber \\
\hspace{-0.25cm} && \hspace{-0.25cm}
+ \hspace{0.06cm} M_4^{} M_5^{} \left[\sum_{i=1}^3 s_{i4}^{} s_{i5}^{}
s_{i^{\prime} 4}^{} s^{}_{i^{\prime}5} \left(s_{i4}^2 s_{i^{\prime\prime}5}^2
+ s_{i^{\prime\prime}4}^2 s_{i^{\prime}5}^2 - s_{i^{\prime\prime}4}^2
s_{i5}^2 - s_{i^{\prime}4}^2 s_{i^{\prime\prime}5}^2\right)
\sin\left(\alpha_i^{} + \alpha^{}_{i^{\prime}}\right) \right.
\nonumber \\
\hspace{-0.25cm} && \hspace{-0.25cm}
- \left. \sum_{i=1}^3 s_{i4}^2 s_{i5}^2 \left(s_{i^{\prime}4}^2
s_{i^{\prime\prime}5}^2 - s_{i^{\prime\prime}4}^2 s_{i^{\prime}5}^2\right)
\sin 2\alpha_i^{} \right] \Biggr\} \; ,
\label{4.8}
\end{eqnarray}
where $i$, $i^{\prime}$ and $i^{\prime\prime}$ run cyclically over 1, 2 and 3.
As discussed in Ref.~\cite{Xing:2024xwb}, ${\cal J}_{\nu}^{}$ in Eq.~(\ref{4.8})
is composed of the sine functions of 9 different phase combinations.
In comparison, the results of $m_i^2$, $H_{\alpha\alpha}^{}$ and
$\left(H^2\right)_{\alpha\alpha}$ (for $i = 1, 2, 3$ and $\alpha = e, \mu, \tau$)
consist of the following phase combinations:
\begin{eqnarray}
\begin{array}{l}
\cos 2\alpha_1^{}\; , ~\cos 2\alpha_2^{} \; ,
~\cos 2\alpha_3^{} \; ; ~
\cos\left(\alpha_1^{} \pm \alpha_2^{}\right) \; , ~
\cos\left(\alpha_2^{} \pm \alpha_3^{}\right) \; , ~
\cos\left(\alpha_3^{} \pm \alpha_1^{}\right) \; ; ~
\nonumber \\
\cos4 \alpha_1^{} \; , ~\cos4\alpha_2^{} \; , ~\cos4 \alpha_3^{} \; ;
~\cos\left(2\alpha_1^{} + \alpha_2^{} + \alpha_3^{}\right) \; ,
~\cos\left(2\alpha_2^{} + \alpha_3^{} + \alpha_1^{}\right) \; ,
~\cos\left(2\alpha_3^{} + \alpha_1^{} + \alpha_2^{}\right) \; ;
\nonumber \\
\cos\left(3\alpha_1^{} + \alpha_2^{}\right) \; ,
~\cos\left(3\alpha_1^{} + \alpha_3^{}\right) \; ,
~\cos\left(3\alpha_2^{} + \alpha_3^{}\right) \; ,
~\cos\left(3\alpha_2^{} + \alpha_1^{}\right) \; ,
~\cos\left(3\alpha_3^{} + \alpha_1^{}\right) \; ,
~\cos\left(3\alpha_3^{} + \alpha_2^{}\right) \; ;
\nonumber\\
\cos2\left(\alpha_1^{} + \alpha_2^{}\right) \; ,
~\cos2\left(\alpha_2^{} + \alpha_3^{}\right) \; ,
~\cos2\left(\alpha_3^{} + \alpha_1^{}\right) \; .
\end{array}
\end{eqnarray}
The above analytical results show that it should be much easier to test
the minimal seesaw scenario with the help of neutrino oscillations.

The analytical expressions for some non-oscillation quantities of massive
neutrinos, such as the sum of three light neutrino masses $\Sigma^{}_\nu$,
the effective mass of a beta decay $\langle m\rangle^{}_e$ and the effective
mass of a neutrinoless double-beta decay $\langle m\rangle^{}_{0\nu2\beta}$,
can also be simplified in the minimal seesaw framework. Therefore, whether
this benchmark mechanism of neutrino mass generation is able to survive the
experimental tests will hopefully be seen in the near future.

\section{Summary and remarks}

So far hundreds of neutrino mass models have been proposed in the literature,
but it remains to be seen which one may finally prove to be true. From the
theoretical point of view, we expect that a most promising model of this
kind should represent a most natural extension of the SM with high gains
and low costs. The canonical seesaw mechanism, which is fully consistent
with the spirit of Weinberg's SMEFT and capable of simultaneously explaining
two big puzzles --- the origin of tiny neutrino masses and the origin of
cosmic matter-antimatter asymmetry, turns out to be the most favorite
candidate in this respect. Then we are left with an important question:
how can the canonical seesaw mechanism be tested in the upcoming precision
measurement era of neutrino physics?

To answer this question, one needs to establish a model-independent bridge
between the original seesaw flavor parameters and all the possibly
observable quantities in particle physics and cosmology. In the present
work we have {\it for the first time} derived the general and explicit
analytical expressions of the 6 neutrino oscillation parameters in terms
of the 18 seesaw flavor parameters without any special assumptions. We
hope that our results will pave the way for testing the seesaw mechanism
at low energies, especially in a variety of neutrino oscillation experiments.

Of course, the naturalness of the canonical seesaw mechanism requires the
mass scale of three heavy Majorana neutrinos to be far above the electroweak
scale. This requirement implies that non-unitarity of the $3\times 3$ PMNS
matrix $U$ must be highly suppressed, as can be seen from the strongly
suppressed active-sterile flavor mixing angles that we have illustrated. On the
other hand, it implies a gap between the flavor parameters of three light
Majorana neutrinos at the seesaw scale and their counterparts that are
observable at the electroweak scale. It is well known that the
renormalization-group equations (RGEs) can help fill the gap, and such quantum
corrections are found to be small in most cases within the seesaw-extended SM
framework~\cite{Ohlsson:2013xva}. That is why we have left aside the RGE
effects in the present work, but they can easily be taken into account when 
necessary, in particular when the input neutrino oscillation data or 
non-oscillation data are accurate and precise enough.

Let us remark that the 9 active-sterile flavor mixing angles of $R$ characterize
the strengths of Yukawa interactions for massive neutrinos, and hence most of
them are not allowed to vanish although some of them might be vanishingly 
small. The reason is simply that $R = {\bf 0}$ would lead to the collapse of 
the seesaw mechanism, as it would result in 
$U D^{}_\nu U^T = - R D^{}_N R^T = {\bf 0}$. So the
largeness of heavy Majorana neutrino masses help compensate the smallness 
of active-sterile flavor mixing, in order to generate tiny masses but significant
flavor mixing effects for the three active neutrinos. This is just the keystone
of ``seesaw". Further constraints on the texture of $R$ via neutrino oscillations 
will allow us to explore the tiny Yukawa interactions between the Higgs and 
neutrino fields. 

A global analysis of all the available experimental data to constrain the
parameter space of the canonical seesaw mechanism is highly nontrivial and
beyond the scope of our paper. In other words, this work is mainly
intended to present the general and explicit analytical results for all
the neutrino oscillation parameters in terms of the original seesaw flavor
parameters. We are going to carry out a comprehensive numerical analysis 
of this kind elsewhere.

\section*{Acknowledgements}

The work of Z.Z.X. is supported in part by the National Natural Science
Foundation of China under grant No. 12075254, and the work of J.Y.Z.
is partly supported by the the National Key Research and Development Program
of China under grant No. 2021YFA1601300, by the National Natural Science
Foundation of China under grant No. 12222512, and by the CAS Pioneer Hundred
Talent Program.

\newpage

\renewcommand{\theequation}{\thesection.\arabic{equation}}
\section*{Appendix}
\appendix

\setcounter{equation}{0}
\section{The expressions of $H^{}_{\alpha\alpha}$ and $\left(H^2\right)_{\alpha\alpha}$}
\label{A}

The more specific analytical expressions of $H_{\alpha\alpha}^{}$ (for
$\alpha = e, \mu, \tau$) defined in Eq.~(\ref{3.2}) are listed below:
\begin{eqnarray}
H_{ee}^{} \hspace{-0.25cm} & = & \hspace{-0.25cm}
s_{14}^2 I_{44}^{} M_4^2 + s_{15}^2 I_{55}^{} M_5^2 +
s_{16}^2 I_{66}^{} M_6^2 + 2 M_4^{} M_5^{} s_{14}^{} s_{15}^{} \left[
\sum_{i=1}^3 s_{i4}^{} s_{i5}^{} \cos\left(\alpha_i^{} + \alpha_1^{}\right)\right]
\nonumber \\
\hspace{-0.25cm} && \hspace{-0.25cm}
+ \hspace{0.06cm} 2 M_5^{} M_6^{} s_{15}^{} s_{16}^{} \left[
\sum_{i=1}^3 s_{i5}^{} s_{i6}^{} \cos\left(\beta_i^{} + \beta_1^{}\right)\right]
+ 2 M_4^{} M_6^{} s_{14}^{} s_{16}^{} \left[
\sum_{i=1}^3 s_{i4}^{} s_{i6}^{} \cos\left(\gamma_i^{} + \gamma_1^{}\right)\right] \; ,
\nonumber \\
H_{\mu\mu}^{} \hspace{-0.25cm} & = & \hspace{-0.25cm}
s_{24}^2 I_{44}^{} M_4^2 + s_{25}^2 I_{55}^{} M_5^2 +
s_{26}^2 I_{66}^{} M_6^2 + 2 M_4^{} M_5^{} s_{24}^{} s_{25}^{} \left[
\sum_{i=1}^3 s_{i4}^{} s_{i5}^{} \cos(\alpha_i^{} + \alpha_2^{})\right]
\nonumber \\
\hspace{-0.25cm} && \hspace{-0.25cm}
+ \hspace{0.06cm} 2 M_5^{} M_6^{} s_{25}^{} s_{26}^{} \left[
\sum_{i=1}^3 s_{i5}^{} s_{i6}^{} \cos\left(\beta_i^{} + \beta_2^{}\right)\right]
+ 2 M_4^{} M_6^{} s_{24}^{} s_{26}^{} \left[
\sum_{i=1}^3 s_{i4}^{} s_{i6}^{} \cos\left(\gamma_i^{} + \gamma_2^{}\right)\right] \; ,
\nonumber \\
H_{\tau\tau}^{} \hspace{-0.25cm} & = & \hspace{-0.25cm}
s_{34}^2 I_{44}^{} M_4^2 + s_{35}^2 I_{55}^{} M_5^2 +
s_{36}^2 I_{66}^{} M_6^2 + 2 M_4^{} M_5^{} s_{34}^{} s_{35}^{} \left[
\sum_{i=1}^3 s_{i4}^{} s_{i5}^{} \cos\left(\alpha_i^{} + \alpha_3^{}\right)\right]
\nonumber \\
\hspace{-0.25cm} && \hspace{-0.25cm}
+ \hspace{0.06cm} 2 M_5^{} M_6^{} s_{35}^{} s_{36}^{} \left[
\sum_{i=1}^3 s_{i5}^{} s_{i6}^{} \cos(\beta_i^{} + \beta_3^{})\right]
+ 2 M_4^{} M_6^{} s_{34}^{} s_{36}^{} \left[
\sum_{i=1}^3 s_{i4}^{} s_{i6}^{} \cos\left(\gamma_i^{} + \gamma_3^{}\right)\right] \; .
\hspace{0.9cm}
\label{A.1}
\end{eqnarray}
In comparison, the analytical expressions of $\left(H^2\right)_{\alpha\alpha}^{}$ (for
$\alpha = e, \mu, \tau$) defined in Eq.~(\ref{3.2}) can be decomposed into the following
15 terms:
\begin{eqnarray}
\left(H^2\right)_{\alpha\alpha} \hspace{-0.25cm} & = & \hspace{-0.25cm}
T^{\alpha\alpha}_{400} M_4^4 +  T^{\alpha\alpha}_{040} M_5^4
+ T^{\alpha\alpha}_{004} M_6^4
+ T^{\alpha\alpha}_{112} M_4^{} M_5^{} M_6^{2} + T^{\alpha\alpha}_{121} M_4^{} M_5^{2}
M_6^{} + T^{\alpha\alpha}_{112} M_4^{} M_5^{} M_6^{2} \hspace{0.9cm}
\nonumber \\
\hspace{-0.25cm} && \hspace{-0.25cm}
+ \hspace{0.06cm} T^{\alpha\alpha}_{130}M_4^{} M_5^3 + T^{\alpha\alpha}_{103}M_4^{} M_6^3
+ T^{\alpha\alpha}_{013}M_5^{} M_6^3 + T^{\alpha\alpha}_{310}M_4^{3} M_5^{} +
T^{\alpha\alpha}_{301}M_4^{3} M_6^{}
\nonumber \\
\hspace{-0.25cm} && \hspace{-0.25cm}
+ \hspace{0.06cm} T^{\alpha\alpha}_{031}M_5^{3} M_6^{}
+ T^{\alpha\alpha}_{220}M_4^{2} M_5^{2}
+ T^{\alpha\alpha}_{022}M_5^{2} M_6^{2}
+ T^{\alpha\alpha}_{202} M_4^{2} M_6^{2} \; ,
\label{A.2}
\end{eqnarray}
where each $T^{\alpha\alpha}_{pqt} $ represents the coefficient of the term
proportional to $M_4^{p} M_5^{q} M_{6}^{t}$ with $p$, $q$ and $t$ being the
simple integers. To be more explicit, we obtain the expressions
\begin{eqnarray}
T_{400}^{ee} \hspace{-0.25cm} & = & \hspace{-0.25cm}
I_{44}^3 s_{14}^2 \; , \quad
T_{040}^{ee} = I_{55}^3 s_{15}^2 \; , \quad
T_{004}^{ee} = I_{66}^3 s_{16}^2 \; ,
\nonumber \\
T_{112}^{ee} \hspace{-0.25cm} & = & \hspace{-0.25cm}
2 I_{66}^{} \Bigg\{ s_{14}^{} s_{15}^{}
\left[\sum^3_{i=1} \sum^3_{i^{\prime}_{}=1} s_{i5}^{} s_{i6}^{} s_{i^{\prime}_{}4}^{}
s_{i^{\prime}_{}6}^{} \cos\left(\beta_i^{} + \gamma_{i^{\prime}_{}}^{} - \alpha_1^{}\right)
\right]
\nonumber \\
\hspace{-0.25cm} && \hspace{-0.25cm}
+ \hspace{0.06cm} s_{15}^{} s_{16}^{}
\left[\sum^3_{i=1} \sum^3_{i^{\prime}_{}=1} s_{i4}^{} s_{i5}^{} s_{i^{\prime}_{}4}^{}
s_{i^{\prime}_{}6}^{} \cos\left(\alpha_i^{} -  \gamma_{i^{\prime}_{}}^{} - \beta_1^{}\right)
\right]
\nonumber \\
\hspace{-0.25cm} && \hspace{-0.25cm}
+ \hspace{0.06cm} s_{14}^{} s_{16}^{}
\left[\sum^3_{i=1} \sum^3_{i^{\prime}_{}=1} s_{i4}^{} s_{i5}^{} s_{i^{\prime}_{}5}^{}
s_{i^{\prime}_{}6}^{} \cos\left(\alpha_i^{} - \beta_{i^{\prime}_{}}^{} - \gamma_1^{}\right)
\right] \Bigg\}
\nonumber \\
\hspace{-0.25cm} && \hspace{-0.25cm}
+ \hspace{0.06cm} 2 s_{14}^{} s_{16}^{} \Biggl[ \sum^3_{i=1}
\sum^3_{i^{\prime} =1} \sum^3_{i^{\prime\prime}=1}
s_{i5}^{} s_{i6}^{} s_{i^{\prime}5}^{} s_{i^{\prime}6}^{} s_{i^{\prime\prime}4}^{}
s_{i^{\prime\prime}6}^{} \cos\left(\beta_i^{} +\beta_{i^{\prime}}^{} -
\gamma_{i^{\prime\prime}}^{} - \gamma_1^{}\right) \Biggr]
\nonumber \\
\hspace{-0.25cm} && \hspace{-0.25cm}
+ \hspace{0.06cm} 2 s_{15}^{} s_{16}^{} \Biggl[ \sum^3_{i=1}
\sum^3_{i^{\prime} =1} \sum^3_{i^{\prime\prime}=1}
s_{i4}^{} s_{i6}^{} s_{i^{\prime}4}^{} s_{i^{\prime}6}^{} s_{i^{\prime\prime}5}^{}
s_{i^{\prime\prime}6}^{} \cos\left(\gamma_i^{} + \gamma_{i^{\prime}}^{} -
\beta_{i^{\prime\prime}}^{} - \beta_1^{}\right) \Biggr]
\nonumber \\
\hspace{-0.25cm} && \hspace{-0.25cm}
+ \hspace{0.06cm} 2 s_{16}^2 \Biggl[ \sum^3_{i=1}
\sum^3_{i^{\prime} =1} \sum^3_{i^{\prime\prime}=1}
s_{i4}^{} s_{i5}^{} s_{i^{\prime}5}^{} s_{i^{\prime}6}^{} s_{i^{\prime\prime}4}^{}
s_{i^{\prime\prime}6}^{} \cos\left(\alpha_i^{} - \beta_{i^{\prime}}^{} -
\gamma_{i^{\prime\prime}}^{}\right)\Biggr] \; ,
\nonumber \\
T_{130}^{ee} \hspace{-0.25cm} & = & \hspace{-0.25cm}
2 I_{55}^2 s_{14}^{} s_{15}^{} \left[ \sum_{i=1}^3
s_{i4}^{} s_{i5}^{} \cos\left(\alpha_i^{} +\alpha_1^{}\right)\right]
+ 2 I_{55}^{} s_{15}^2
\left[ \sum^3_{i=1} \sum^3_{i^{\prime}=1} s_{i4}^{} s_{i5}^{} s_{i^{\prime}4}^{}
s_{i^{\prime}5}^{}\cos\left(\alpha_i^{} + \alpha_{i^{\prime}}^{}\right)\right] \; ,
\nonumber \\
T_{220}^{ee} \hspace{-0.25cm} & = & \hspace{-0.25cm}
\left(I_{44}^{} s_{15}^2 + I_{55}^{} s_{14}^2\right)
\left[ \sum^3_{i=1} \sum^3_{i^{\prime}=1} s_{i4}^{} s_{i5}^{} s_{i^{\prime}4}^{}
s_{i^{\prime}5}^{}\cos\left(\alpha_i^{} -\alpha_{i^{\prime}}^{}\right)\right]
+ 2 s_{14}^{} s_{15}^{} I_{44}^{} I _{55}^{} \left[ \sum_{i=1}^3
s_{i4}^{} s_{i5}^{} \cos\left(\alpha_i^{} -\alpha_1^{}\right)\right]
\nonumber \\
\hspace{-0.25cm} && \hspace{-0.25cm}
+ \hspace{0.06cm} 2 s_{14}^{} s_{15}^{} \Biggl[ \sum^3_{i=1}
\sum^3_{i^{\prime} =1} \sum^3_{i^{\prime\prime}=1}
s_{i4}^{} s_{i5}^{} s_{i^{\prime}4}^{} s_{i^{\prime}5}^{} s_{i^{\prime\prime}4}^{}
s_{i^{\prime\prime}5}^{} \cos\left(\alpha_i^{} + \alpha_{i^{\prime}}^{} +
\alpha_{i^{\prime\prime}}^{} + \alpha_1^{}\right) \Biggr] \; ;
\label{A.3}
\end{eqnarray}
\begin{eqnarray}
T_{400}^{\mu\mu} \hspace{-0.25cm} & = & \hspace{-0.25cm}
I_{44}^3 s_{24}^2 \; , \quad
T_{040}^{\mu\mu} = I_{55}^3 s_{25}^2 \; , \quad
T_{004}^{\mu\mu} = I_{66}^3 s_{26}^2 \; ,
\nonumber \\
T_{112}^{\mu\mu} \hspace{-0.25cm} & = & \hspace{-0.25cm}
2 I_{66}^{} \Bigg\{ s_{24}^{} s_{25}^{}
\left[\sum_{i, i^{\prime}_{}=1}^3 s_{i5}^{} s_{i6}^{} s_{i^{\prime}_{}4}^{}
s_{i^{\prime}_{}6}^{} \cos\left(\beta_i^{} + \gamma_{i^{\prime}_{}}^{} - \alpha_2^{}\right)
\right]
\nonumber \\
\hspace{-0.25cm} && \hspace{-0.25cm}
+ \hspace{0.06cm} s_{25}^{} s_{26}^{}
\left[\sum^3_{i=1} \sum^3_{i^{\prime} =1} s_{i4}^{} s_{i5}^{}
s_{i^{\prime}_{}4}^{} s_{i^{\prime}_{}6}^{}	\cos\left(\alpha_i^{} - \gamma_{i^{\prime}_{}}^{}
- \beta_2^{}\right)\right]
\nonumber \\
\hspace{-0.25cm} && \hspace{-0.25cm}
+ \hspace{0.06cm} s_{24}^{} s_{26}^{}
\left[\sum^3_{i=1} \sum^3_{i^{\prime} =1} s_{i4}^{} s_{i5}^{} s_{i^{\prime}_{}5}^{}
s_{i^{\prime}_{}6}^{} \cos\left(\alpha_i^{} - \beta_{i^{\prime}_{}}^{} - \gamma_2^{}\right)
\right] \Bigg\}
\nonumber \\
\hspace{-0.25cm} && \hspace{-0.25cm}
+ \hspace{0.06cm} 2 s_{24}^{} s_{26}^{} \Biggl[ \sum^3_{i=1}
\sum^3_{i^{\prime} =1} \sum^3_{i^{\prime\prime}=1}
s_{i5}^{} s_{i6}^{} s_{i^{\prime}5}^{} s_{i^{\prime}6}^{} s_{i^{\prime\prime}4}^{}
s_{i^{\prime\prime}6}^{} \cos\left(\beta_i^{} +\beta_{i^{\prime}}^{} -
\gamma_{i^{\prime\prime}}^{} - \gamma_2^{}\right) \Biggr]
\nonumber \\
\hspace{-0.25cm} && \hspace{-0.25cm}
+ \hspace{0.06cm} 2 s_{25}^{} s_{26}^{} \Biggl[ \sum^3_{i=1}
\sum^3_{i^{\prime} =1} \sum^3_{i^{\prime\prime}=1}
s_{i4}^{} s_{i6}^{} s_{i^{\prime}4}^{} s_{i^{\prime}6}^{} s_{i^{\prime\prime}5}^{}
s_{i^{\prime\prime}6}^{} \cos\left(\gamma_i^{} + \gamma_{i^{\prime}}^{} -
\beta_{i^{\prime\prime}}^{} - \beta_2^{}\right) \Biggr]
\nonumber \\
\hspace{-0.25cm} && \hspace{-0.25cm}
+ \hspace{0.06cm} 2 s_{26}^2 \Biggl[ \sum^3_{i=1} \sum^3_{i^{\prime} =1}
\sum^3_{i^{\prime\prime}=1} s_{i4}^{} s_{i5}^{} s_{i^{\prime}5}^{} s_{i^{\prime}6}^{}
s_{i^{\prime\prime}4}^{} s_{i^{\prime\prime}6}^{} \cos\left(\alpha_i^{} -
\beta_{i^{\prime}}^{} -	\gamma_{i^{\prime\prime}}^{}\right) \Biggr] \; ,
\nonumber \\
T_{130}^{\mu\mu} \hspace{-0.25cm} & = & \hspace{-0.25cm}
2 I_{55}^2 s_{24}^{} s_{25}^{} \left[ \sum_{i=1}^3
s_{i4}^{} s_{i5}^{} \cos\left(\alpha_i^{} +\alpha_2^{}\right)\right] +
2 I_{55}^{} s_{25}^2 \left[ \sum^3_{i=1} \sum^3_{i^{\prime}=1}
s_{i4}^{} s_{i5}^{} s_{i^{\prime}4}^{} s_{i^{\prime}5}^{}\cos\left(\alpha_i^{}
+ \alpha_{i^{\prime}}^{}\right)\right] \; ,
\nonumber \\
T_{220}^{\mu\mu} \hspace{-0.25cm} & = & \hspace{-0.25cm}
\left(I_{44}^{} s_{25}^2 + I_{55}^{} s_{24}^2\right)
\left[ \sum^3_{i=1} \sum^3_{i^{\prime}=1} s_{i4}^{} s_{i5}^{} s_{i^{\prime}4}^{}
s_{i^{\prime}5}^{}\cos\left(\alpha_i^{} -\alpha_{i^{\prime}}^{}\right)\right]
+ 2 s_{24}^{} s_{25}^{} I_{44}^{} I _{55}^{} \left[ \sum_{i=1}^3
s_{i4}^{} s_{i5}^{} \cos\left(\alpha_i^{} - \alpha_2^{}\right)\right]
\nonumber \\
\hspace{-0.25cm} && \hspace{-0.25cm}
+ \hspace{0.06cm} 2 s_{24}^{} s_{25}^{} \Biggl[ \sum^3_{i=1}
\sum^3_{i^{\prime} =1} \sum^3_{i^{\prime\prime}=1}
s_{i4}^{} s_{i5}^{} s_{i^{\prime}4}^{} s_{i^{\prime}5}^{} s_{i^{\prime\prime}4}^{}
s_{i^{\prime\prime}5}^{} \cos\left(\alpha_i^{} + \alpha_{i^{\prime}}^{} +
\alpha_{i^{\prime\prime}}^{} + \alpha_2^{}\right) \Biggr] \; ;
\label{A.4}
\end{eqnarray}
\begin{eqnarray}
T_{400}^{\tau\tau} \hspace{-0.25cm} & = & \hspace{-0.25cm}
I_{44}^3 s_{34}^2 \; , \quad
T_{040}^{\tau\tau} = I_{55}^3 s_{35}^2 \; , \quad
T_{004}^{\tau\tau} = I_{66}^3 s_{36}^2 \; ,
\nonumber \\
T_{112}^{\tau\tau} \hspace{-0.25cm} & = & \hspace{-0.25cm}
2 I_{66}^{} \Bigg\{ s_{34}^{} s_{35}^{}
\left[\sum^3_{i=1} \sum^3_{i^{\prime}_{}=1} s_{i5}^{} s_{i6}^{}  s_{i^{\prime}_{}4}^{}
s_{i^{\prime}_{}6}^{} \cos\left(\beta_i^{} + \gamma_{i^{\prime}_{}}^{} - \alpha_3^{}\right)
\right]
\nonumber \\
\hspace{-0.25cm} && \hspace{-0.25cm}
+ \hspace{0.06cm} s_{35}^{} s_{36}^{}
\left[\sum^3_{i=1} \sum^3_{i^{\prime}_{}=1} s_{i4}^{} s_{i5}^{} s_{i^{\prime}_{}4}^{}
s_{i^{\prime}_{}6}^{} \cos\left(\alpha_i^{} - \gamma_{i^{\prime}_{}}^{} - \beta_3^{}\right)
\right]
\nonumber \\
\hspace{-0.25cm} && \hspace{-0.25cm}
+ \hspace{0.06cm} s_{34}^{} s_{36}^{}
\left[\sum_{i, i^{\prime}_{}=1}^3 s_{i4}^{} s_{i5}^{} s_{i^{\prime}_{}5}^{}
s_{i^{\prime}_{}6}^{} \cos\left(\alpha_i^{} -  \beta_{i^{\prime}_{}}^{} - \gamma_3^{}\right)
\right] \Bigg\}
\nonumber \\
\hspace{-0.25cm} && \hspace{-0.25cm}
+ \hspace{0.06cm} 2 s_{34}^{} s_{36}^{} \Biggl[ \sum^3_{i=1} \sum^3_{i^{\prime} =1}
\sum^3_{i^{\prime\prime}=1} s_{i5}^{} s_{i6}^{} s_{i^{\prime}5}^{} s_{i^{\prime}6}^{}
s_{i^{\prime\prime}4}^{} s_{i^{\prime\prime}6}^{} \cos\left(\beta_i^{} +
\beta_{i^{\prime}}^{} -	\gamma_{i^{\prime\prime}}^{} - \gamma_3^{}\right)
\Biggr]
\nonumber \\
\hspace{-0.25cm} && \hspace{-0.25cm}
+ \hspace{0.06cm} 2 s_{35}^{} s_{36}^{} \Biggl[ \sum^3_{i=1}
\sum^3_{i^{\prime} =1} \sum^3_{i^{\prime\prime}=1}
s_{i4}^{} s_{i6}^{} s_{i^{\prime}4}^{} s_{i^{\prime}6}^{} s_{i^{\prime\prime}5}^{}
s_{i^{\prime\prime}6}^{} \cos\left(\gamma_i^{} + \gamma_{i^{\prime}}^{} -
\beta_{i^{\prime\prime}}^{} - \beta_3^{}\right) \Biggr]
\nonumber \\
\hspace{-0.25cm} && \hspace{-0.25cm}
+ \hspace{0.06cm} 2 s_{36}^2 \Biggl[ \sum^3_{i=1} \sum^3_{i^{\prime} =1}
\sum^3_{i^{\prime\prime}=1} s_{i4}^{} s_{i5}^{} s_{i^{\prime}5}^{} s_{i^{\prime}6}^{}
s_{i^{\prime\prime}4}^{} s_{i^{\prime\prime}6}^{} \cos\left(\alpha_i^{} -
\beta_{i^{\prime}}^{} -	\gamma_{i^{\prime\prime}}^{}\right) \Biggr] \; ,
\nonumber \\
T_{130}^{\tau\tau} \hspace{-0.25cm} & = & \hspace{-0.25cm}
2 I_{55}^2 s_{34}^{} s_{35}^{} \left[ \sum_{i=1}^3
s_{i4}^{} s_{i5}^{} \cos\left(\alpha_i^{} +\alpha_3^{}\right)\right] +
2 I_{55}^{} s_{35}^2 \left[ \sum^3_{i=1} \sum^3_{i^{\prime}=1} s_{i4}^{}
s_{i5}^{} s_{i^{\prime}4}^{} s_{i^{\prime}5}^{}\cos\left(\alpha_i^{} +
\alpha_{i^{\prime}}^{}\right)\right] \; ,
\nonumber\\
T_{220}^{\tau\tau} \hspace{-0.25cm} & = & \hspace{-0.25cm}
\left(I_{44}^{} s_{35}^2 + I_{55}^{} s_{34}^2\right)
\left[ \sum^3_{i=1} \sum^3_{i^{\prime}=1} s_{i4}^{} s_{i5}^{} s_{i^{\prime}4}^{}
s_{i^{\prime}5}^{}\cos\left(\alpha_i^{} - \alpha_{i^{\prime}}^{}\right)\right]
+ 2 s_{34}^{} s_{35}^{} I_{44}^{} I _{55}^{} \left[ \sum_{i=1}^3
s_{i4}^{} s_{i5}^{} \cos\left(\alpha_i^{} - \alpha_3^{}\right)\right]
\nonumber \\
\hspace{-0.25cm} && \hspace{-0.25cm}
+ \hspace{0.06cm} 2 s_{34}^{} s_{35}^{} \Biggl[ \sum^3_{i=1}
\sum^3_{i^{\prime} =1} \sum^3_{i^{\prime\prime}=1}
s_{i4}^{} s_{i5}^{} s_{i^{\prime}4}^{} s_{i^{\prime}5}^{} s_{i^{\prime\prime}4}^{}
s_{i^{\prime\prime}5}^{} \cos\left(\alpha_i^{} + \alpha_{i^{\prime}}^{} +
\alpha_{i^{\prime\prime}}^{} + \alpha_3^{}\right) \Biggr] \; ;
\label{A.5}
\end{eqnarray}
and the other coefficients can simply be achieved from Eqs.~(\ref{A.3}),
(\ref{A.4}) and (\ref{A.5}) by making the following subscript and phase
replacements:
\begin{eqnarray}
T_{121}^{\alpha\alpha} \hspace{-0.25cm} & = & \hspace{-0.25cm}
T_{112}^{\alpha\alpha} \left\{ \left(4, 5, 6\right) \to
\left(4, 6, 5\right) \; ; ~\left(\alpha_i^{}, \beta_{i}^{}, \gamma_i^{}\right)
\to -\left(\gamma_i^{}, \beta_i^{}, \alpha_i^{}\right)\right\} \; ,
\nonumber \\
T_{211}^{\alpha\alpha} \hspace{-0.25cm} & = & \hspace{-0.25cm}
T_{112}^{\alpha\alpha} \left\{ \left(4, 5, 6\right) \to \left(6, 5, 4\right) \; ;
~\left(\alpha_i^{}, \beta_{i}^{}, \gamma_i^{}\right) \to
-\left(\beta_i^{}, \alpha_i^{}, \gamma_i^{}\right)\right\} \; ,
\nonumber \\
T_{103}^{\alpha\alpha} \hspace{-0.25cm} & = & \hspace{-0.25cm}
T_{130}^{\alpha\alpha} \left\{ \left(4, 5, 6\right) \to \left(4, 6, 5\right) \; ;
~\left(\alpha_i^{}, \beta_{i}^{}, \gamma_i^{}\right) \to
-\left(\gamma_i^{}, \beta_i^{}, \alpha_i^{}\right) \right\} \; ,
\nonumber \\
T_{013}^{ee} \hspace{-0.25cm} & = & \hspace{-0.25cm}
T_{130}^{ee} \left\{ \left(4, 5, 6\right) \to \left(5, 6, 4\right) \; ;
~\left(\alpha_i^{}, \beta_{i}^{}, \gamma_i^{}\right) \to +\left(\beta_i^{},
\gamma_i^{}, \alpha_i^{}\right) \right\} \; ,
\nonumber \\
T_{310}^{\alpha\alpha} \hspace{-0.25cm} & = & \hspace{-0.25cm}
T_{130}^{\alpha\alpha} \left\{ \left(4, 5, 6\right) \to \left(5, 4, 6\right) \; ;
~\left(\alpha_i^{}, \beta_{i}^{}, \gamma_i^{}\right) \to
-\left(\alpha_i^{}, \gamma_i^{}, \beta_i^{}\right) \right\} \; ,
\nonumber \\
T_{301}^{\alpha\alpha} \hspace{-0.25cm} & = & \hspace{-0.25cm}
T_{130}^{\alpha\alpha} \left\{ \left(4, 5, 6\right) \to \left(6, 4, 5\right) \; ;
~\left(\alpha_i^{}, \beta_{i}^{}, \gamma_i^{}\right)\to
+\left(\gamma_i^{}, \alpha_i^{}, \beta_i^{}\right) \right\} \; ,
\nonumber \\
T_{031}^{\alpha\alpha} \hspace{-0.25cm} & = & \hspace{-0.25cm}
T_{130}^{\alpha\alpha} \left\{ \left(4, 5, 6\right) \to \left(6, 5, 4\right) \; ;
~\left(\alpha_i^{}, \beta_{i}^{}, \gamma_i^{}\right) \to
-\left(\beta_i^{}, \alpha_i^{}, \gamma_i^{}\right) \right\} \; ,
\nonumber \\
T_{202}^{\alpha\alpha} \hspace{-0.25cm} & = & \hspace{-0.25cm}
T_{220}^{\alpha\alpha} \left\{ \left(4, 5, 6\right) \to \left(4, 6, 5\right) \; ;
~\left(\alpha_i^{}, \beta_{i}^{}, \gamma_i^{}\right) \to
-\left(\gamma_i^{}, \beta_i^{}, \alpha_i^{}\right) \right\} \; ,
\nonumber \\
T_{022}^{\alpha\alpha} \hspace{-0.25cm} & = & \hspace{-0.25cm}
T_{220}^{\alpha\alpha} \left\{ \left(4, 5, 6\right) \to \left(5, 6, 4\right) \; ;
~\left(\alpha_i^{}, \beta_{i}^{}, \gamma_i^{}\right) \to
+\left(\beta_i^{}, \gamma_i^{}, \alpha_i^{}\right) \right\} \; , \hspace{0.7cm}
\label{A.6}
\end{eqnarray}
where the Greek superscript ``$\alpha\alpha$" runs over $ee$, $\mu\mu$ and
$\tau\tau$. It is easy to see that $H_{\mu\mu}^{}$ can be read off from
$H_{ee}^{}$ by doing the replacements $s_{14}^{} \leftrightarrow s^{}_{24}$,
$s_{15}^{} \leftrightarrow s^{}_{25}$, $s_{16}^{} \leftrightarrow s^{}_{26}$, $\alpha_1^{}
\leftrightarrow \alpha_2^{}$, $\beta_1^{} \leftrightarrow \beta_2^{}$ and $\gamma_1^{} \leftrightarrow \gamma_2^{}$.
We find that $\left(H^2\right)_{\mu\mu}^{}$ can be obtained from
$(H^2)_{ee}^{}$ with the same subscript replacements.
Similarly, $H_{\tau\tau}^{}$ and $\left(H^2\right)_{\tau\tau}^{}$ can easily be
read off from $H_{ee}^{}$ and $(H^2)_{ee}^{}$ by doing the
replacements $s_{14}^{} \leftrightarrow s^{}_{34}$, $s_{15}^{} \leftrightarrow s^{}_{35}$,
$s_{16}^{} \leftrightarrow s^{}_{36}$, $\alpha_1^{} \leftrightarrow \alpha_3^{}$,
$\beta_1^{} \leftrightarrow \beta_3^{}$ and $\gamma_1^{} \leftrightarrow \gamma_3^{}$.

It is worth pointing out that the phase combinations that appear in the three
active flavor mixing angles and the effective Dirac CP-violating phase
shown in Eqs.~(\ref{3.17}) and (\ref{3.18}) include
\begin{eqnarray}
\begin{array}{l}
\cos\left(\alpha_i^{} \pm \alpha_{i^{\prime}}^{}\right) \; ,
~\cos\left(\beta_i^{} \pm \beta_{i^{\prime}}^{}\right) \; ,
~\cos\left(\gamma_i^{} \pm \gamma_{i^{\prime}}^{}\right) \; ;
~\cos\left(\alpha_i^{} + \beta_{i^{\prime}}^{}\right) \; ,
~\cos\left(\beta_i^{} + \gamma_{i^{\prime}}^{}\right) \; ,
~\cos\left(\gamma_i^{} + \alpha_{i^{\prime}}^{}\right) \; ;
\nonumber \\
\cos\left(\alpha_i^{} - \beta_{i^{\prime}} + \gamma_{i^{\prime\prime}}\right) \; ,
~\cos\left(\alpha_i^{} + \beta_{i^{\prime}} - \gamma_{i^{\prime\prime}}\right) \; ,
~\cos\left(\beta_i^{} + \gamma_{i^{\prime}} - \alpha_{i^{\prime\prime}}\right) \; ,
~\cos\left(\alpha_i^{} + \beta_{i^{\prime}} + \gamma_{i^{\prime\prime}}\right) \; ;
\nonumber \\
\cos\left(\alpha_i^{} + \alpha_{i^{\prime}}^{} + \alpha_{i^{\prime\prime}}^{}
+ \alpha_{i^{\prime\prime}}^{}\right) \; ,
~\cos\left(\beta_i^{} + \beta_{i^{\prime}}^{} + \beta_{i^{\prime\prime}}^{}
+ \beta_{i^{\prime\prime\prime}}^{}\right) \; ,
~\cos\left(\gamma_i^{} + \gamma_{i^{\prime}}^{} + \gamma_{i^{\prime\prime}}^{}
+ \gamma_{i^{\prime\prime\prime}}^{}\right) \; ;
\nonumber \\
\cos\left(\alpha_i^{} + \alpha_{i^{\prime}}^{} - \beta_{i^{\prime\prime}}^{} -
\beta_{i^{\prime\prime\prime}}^{}\right) \; ,
~\cos\left(\beta_i^{} + \beta_{i^{\prime}}^{} - \gamma_{i^{\prime\prime}}^{} -
\gamma_{i^{\prime\prime\prime}}^{}\right) \; ,
~\cos\left(\gamma_i^{} + \gamma_{i^{\prime}}^{} - \alpha_{i^{\prime\prime}}^{}
- \alpha_{i^{\prime\prime\prime}}^{}\right) \; ,
\end{array}
\end{eqnarray}
where $i, i^{\prime}, i^{\prime\prime}, i^{\prime\prime\prime} = 1, 2, 3$.
In total, there are 267 different phase combinations as listed below.
\begin{eqnarray}
\begin{array}{l}
\cos \alpha_1^{} \; , ~\cos \alpha_2^{} \; , ~\cos \alpha_3^{} \; ;
~\cos \beta_1^{} \; , ~\cos \beta_2^{} \; , ~\cos \beta_3^{} \; ;
~\cos \gamma_1^{} \; , ~\cos \gamma_2^{} \; , ~\cos \gamma_3^{} \; ;
\nonumber \\
\cos 2\alpha_1^{} \; , ~\cos 2\alpha_2^{} \; , ~\cos 2\alpha_3^{} \; ;
~\cos2 \beta_1^{} \; , ~\cos 2\beta_2^{} \; , ~\cos 2\beta_3^{} \; ;
~\cos 2\gamma_1^{} \; , ~\cos 2\gamma_2^{} \; , ~\cos 2\gamma_3^{} \; ;
\nonumber \\
\cos 4\alpha_1^{} \; , ~\cos 4\alpha_2^{} \; , ~\cos 4\alpha_3^{} \; ;
~\cos4 \beta_1^{} \; , ~\cos 4\beta_2^{} \; , ~\cos 4\beta_3^{} \; ;
~\cos 4\gamma_1^{} \; , ~\cos 4\gamma_2^{} \; , ~\cos 4\gamma_3^{} \; ;
\nonumber \\
\cos \left(\alpha_1^{} \pm \alpha_2^{}\right) \; ,
~\cos \left(\alpha_2^{} \pm \alpha_3^{}\right) \; ,
~\cos\left(\alpha_3^{} \pm \alpha_1^{}\right) \; ;
~\cos \left(\beta_1^{} \pm \beta_2^{}\right) \; ,
~\cos \left(\beta_2^{} \pm \beta_3^{}\right) \; ,
~\cos \left(\beta_3^{} \pm \beta_1^{}\right) \; ;
\nonumber \\
\cos \left(\gamma_1^{} \pm \gamma_2^{}\right) \; ,
~\cos \left(\gamma_2^{} \pm \gamma_3^{}\right) \; ,
~\cos\left(\gamma_3^{} \pm \gamma_1^{}\right ) \; ;
\nonumber \\
\cos \left(\alpha_1^{} + \beta_2^{}\right) \; ,
~\cos \left(\alpha_1^{} +  \beta_3^{}\right) \; ,
~\cos \left(\alpha_2^{} + \beta_1^{}\right) \; ,
~\cos \left(\alpha_2^{} +  \beta_3^{}\right) \; ,
~\cos \left(\alpha_3^{} + \beta_1^{}\right) \; ,
~\cos \left(\alpha_3^{} +  \beta_2^{} \right) \; ;
\nonumber \\
\cos \left(\alpha_1^{} + \gamma_2^{}\right) \; ,
~\cos \left(\alpha_1^{} +  \gamma_3^{}\right) \; ,
~\cos \left(\alpha_2^{} + \gamma_1^{}\right) \; ,
~\cos \left(\alpha_2^{} +  \gamma_3^{}\right) \; ,
~\cos \left(\alpha_3^{} + \gamma_1^{}\right) \; ,
~\cos \left(\alpha_3^{} +  \gamma_2^{}\right) \; ;
\nonumber \\
\cos \left(\beta_1^{} + \gamma_2^{}\right) \; ,
~\cos \left(\beta_1^{} +  \gamma_3^{}\right) \; ,
~\cos \left(\beta_2^{} + \gamma_1^{}\right) \; ,
~\cos \left(\beta_2^{} +  \gamma_3^{} \right) \; ,
~\cos \left(\beta_3^{} + \gamma_1^{}\right) \; ,
~\cos \left(\beta_3^{} +  \gamma_2^{}\right) \; ;
\nonumber \\
\cos 2\left(\alpha_1^{} + \alpha_2^{}\right) \; ,
~\cos 2\left(\alpha_2^{} +  \alpha_3^{}\right) \; ,
~\cos 2\left(\alpha_3^{} + \alpha_1^{}\right) \; ;
~\cos 2\left(\beta_1^{} + \beta_2^{}\right) \; ,
~\cos 2\left(\beta_2^{} +  \beta_3^{}\right) \; ,
~\cos 2\left(\beta_3^{} + \beta_1^{}\right) \; ;
\nonumber \\
\cos 2\left(\gamma_1^{} + \gamma_2^{}\right) \; ,
~\cos 2\left(\gamma_2^{} +  \gamma_3^{}\right) \; ,
~\cos 2\left(\gamma_3^{} + \gamma_1^{}\right) \; ;
\nonumber \\
\cos2\left(\alpha_1^{}-\beta_1^{}\right) \; ,
~\cos2\left(\alpha_1^{}-\beta_2^{}\right) \; ,
~\cos2\left(\alpha_1^{}-\beta_3^{}\right) \; ,
~\cos2\left(\alpha_2^{}-\beta_1^{}\right) \; ,
~\cos2\left(\alpha_2^{}-\beta_2^{}\right) \; ,
~\cos2\left(\alpha_2^{}-\beta_3^{}\right) \; ,
\nonumber \\
\cos2\left(\alpha_3^{}-\beta_1^{}\right) \; ,
~\cos2\left(\alpha_3^{}-\beta_2^{}\right) \; ,
~\cos2\left(\alpha_3^{}-\beta_3^{}\right) \; ;
~\cos2\left(\alpha_1^{}-\gamma_1^{}\right) \; ,
~\cos2\left(\alpha_1^{}-\gamma_2^{}\right) \; ,
~\cos2\left(\alpha_1^{}-\gamma_3^{}\right) \; ,
\nonumber \\
\cos2\left(\alpha_2^{}-\gamma_1^{}\right) \; ,
~\cos2\left(\alpha_2^{}-\gamma_2^{}\right) \; ,
~\cos2\left(\alpha_2^{}-\gamma_3^{}\right) \; ,
~\cos2\left(\alpha_3^{}-\gamma_1^{}\right) \; ,
~\cos2\left(\alpha_3^{}-\gamma_2^{}\right) \; ,
~\cos2\left(\alpha_3^{}-\gamma_3^{}\right) \; ;
\nonumber \\
\cos2\left(\beta_1^{}-\gamma_1^{}\right) \; ,
~\cos2\left(\beta_1^{}-\gamma_2^{}\right) \; ,
~\cos2\left(\beta_1^{}-\gamma_3^{}\right) \; ,
~\cos2\left(\beta_2^{}-\gamma_1^{}\right) \; ,
~\cos2\left(\beta_2^{}-\gamma_2^{}\right) \; ,
~\cos2\left(\beta_2^{}-\gamma_3^{}\right) \; ,
\nonumber \\
\cos2\left(\beta_3^{}-\gamma_1^{}\right) \; ,
~\cos2\left(\beta_3^{}-\gamma_2^{}\right) \; ,
~\cos2\left(\beta_3^{}-\gamma_3^{}\right) \; ;
\nonumber \\
\cos \left(3\alpha_1^{} + \alpha_2^{}\right) \; ,
~\cos \left(3\alpha_1^{} +  \alpha_3^{}\right) \; ,
~\cos \left(3\alpha_2^{} + \alpha_1^{}\right) \; ,
~\cos \left(3\alpha_2^{} + \alpha_3^{}\right) \; ,
~\cos \left(3\alpha_3^{} + \alpha_1^{}\right) \; ,
~\cos \left(3\alpha_3^{} + \alpha_2^{}\right) \; ;
\nonumber \\
\cos \left(3\beta_1^{} + \beta_2^{}\right) \; ,
~\cos \left(3\beta_1^{} +  \beta_3^{}\right) \; ,
~\cos \left(3\beta_2^{} + \beta_1^{}\right) \; ,
~\cos \left(3\beta_2^{} + \beta_3^{}\right) \; ,
~\cos \left(3\beta_3^{} + \beta_1^{}\right) \; ,
~\cos \left(3\beta_3^{} + \beta_2^{}\right) \; ;
\nonumber \\
\cos \left(3\gamma_1^{} + \gamma_2^{}\right) \; ,
~\cos \left(3\gamma_1^{} +  \gamma_3^{}\right) \; ,
~\cos \left(3\gamma_2^{} + \gamma_1^{}\right) \; ,
~\cos \left(3\gamma_2^{} + \gamma_3^{}\right) \; ,
~\cos \left(3\gamma_3^{} + \gamma_1^{}\right) \; ,
~\cos \left(3\gamma_3^{} + \gamma_2^{}\right) \; ;
\nonumber \\
\cos\left(2\alpha_1^{} - \beta_1^{} -\beta_2^{}\right) \; ,
~\cos\left(2\alpha_1^{} - \beta_1^{} -\beta_3^{}\right) \; ,
~\cos\left(2\alpha_1^{} - \beta_2^{} -\beta_3^{}\right) \; ,
~\cos\left(2\alpha_2^{} - \beta_1^{} -\beta_2^{}\right) \; ,
\\
\cos\left(2\alpha_2^{} - \beta_1^{} -\beta_3^{}\right) \; ,
~\cos\left(2\alpha_2^{} - \beta_2^{} -\beta_3^{}\right) \; ,
~\cos\left(2\alpha_3^{} - \beta_1^{} -\beta_2^{}\right) \; ,
~\cos\left(2\alpha_3^{} - \beta_1^{} -\beta_3^{}\right) \; ,
\\
\cos\left(2\alpha_3^{} - \beta_2^{} -\beta_3^{}\right) \; ;
~\cos\left(2\alpha_1^{} - \gamma_1^{} -\gamma_2^{}\right) \; ,
~\cos\left(2\alpha_1^{} - \gamma_1^{} -\gamma_3^{}\right) \; ,
~\cos\left(2\alpha_1^{} - \gamma_2^{} -\gamma_3^{}\right) \; ,
\\
\cos\left(2\alpha_2^{} - \gamma_1^{} -\gamma_2^{}\right) \; ,
~\cos\left(2\alpha_2^{} - \gamma_1^{} -\gamma_3^{}\right) \; ,
~\cos\left(2\alpha_2^{} - \gamma_2^{} -\gamma_3^{}\right) \; ,
~\cos\left(2\alpha_3^{} - \gamma_1^{} -\gamma_2^{}\right) \; ,
\\
\cos\left(2\alpha_3^{} - \gamma_1^{} -\gamma_3^{}\right) \; ,
~\cos\left(2\alpha_3^{} - \gamma_2^{} -\gamma_3^{}\right) \; ;
~\cos\left(2\beta_1^{} - \alpha_1^{} -\alpha_2^{}\right) \; ,
~\cos\left(2\beta_1^{} - \alpha_1^{} -\alpha_3^{}\right) \; ,
\\
\cos\left(2\beta_1^{} - \alpha_2^{} -\alpha_3^{}\right) \; ,
~\cos\left(2\beta_2^{} - \alpha_1^{} -\alpha_2^{}\right) \; ,
~\cos\left(2\beta_2^{} - \alpha_1^{} -\alpha_3^{}\right) \; ,
~\cos\left(2\beta_2^{} - \alpha_2^{} -\alpha_3^{}\right) \; ,
\\
\cos\left(2\beta_3^{} - \alpha_1^{} -\alpha_2^{}\right) \; ,
~\cos\left(2\beta_3^{} - \alpha_1^{} -\alpha_3^{}\right) \; ,
~\cos\left(2\beta_3^{} - \alpha_2^{} -\alpha_3^{}\right) \; ;
~\cos\left(2\beta_1^{} - \gamma_1^{} -\gamma_2^{}\right) \; ,
\\
\cos\left(2\beta_1^{} - \gamma_1^{} -\gamma_3^{}\right) \; ,
~\cos\left(2\beta_1^{} - \gamma_2^{} -\gamma_3^{}\right) \; ,
~\cos\left(2\beta_2^{} - \gamma_1^{} -\gamma_2^{}\right) \; ,
~\cos\left(2\beta_2^{} - \gamma_1^{} -\gamma_3^{}\right) \; ,
\\
\cos\left(2\beta_2^{} - \gamma_2^{} -\gamma_3^{}\right) \; ,
~\cos\left(2\beta_3^{} - \gamma_1^{} -\gamma_2^{}\right) \; ,
~\cos\left(2\beta_3^{} - \gamma_1^{} -\gamma_3^{}\right) \; ,
~\cos\left(2\beta_3^{} - \gamma_2^{} -\gamma_3^{}\right) \; ;
\\
\cos\left(2\gamma_1^{} - \beta_1^{} -\beta_2^{}\right) \; ,
~\cos\left(2\gamma_1^{} - \beta_1^{} -\beta_3^{}\right) \; ,
~\cos\left(2\gamma_1^{} - \beta_2^{} -\beta_3^{}\right) \; ,
~\cos\left(2\gamma_2^{} - \beta_1^{} -\beta_2^{}\right) \; ,
\\
\cos\left(2\gamma_2^{} - \beta_1^{} -\beta_3^{}\right) \; ,
~\cos\left(2\gamma_2^{} - \beta_2^{} -\beta_3^{}\right) \; ,
~\cos\left(2\gamma_3^{} - \beta_1^{} -\beta_2^{}\right) \; ,
~\cos\left(2\gamma_3^{} - \beta_1^{} -\beta_3^{}\right) \; ,
\\
\cos\left(2\gamma_3^{} - \beta_2^{} -\beta_3^{}\right) \; ;
~\cos\left(2\gamma_1^{} - \alpha_1^{} -\alpha_2^{}\right) \; ,
~\cos\left(2\gamma_1^{} - \alpha_1^{} -\alpha_3^{}\right) \; ,
~\cos\left(2\gamma_1^{} - \alpha_2^{} -\alpha_3^{}\right) \; ,
\\
\cos\left(2\gamma_2^{} - \alpha_1^{} -\alpha_2^{}\right) \; ,
~\cos\left(2\gamma_2^{} - \alpha_1^{} -\alpha_3^{}\right) \; ,
~\cos\left(2\gamma_2^{} - \alpha_2^{} -\alpha_3^{}\right) \; ,
~\cos\left(2\gamma_3^{} - \alpha_1^{} -\alpha_2^{}\right) \; ,
\\
\cos\left(2\gamma_3^{} - \alpha_1^{} -\alpha_3^{}\right) \; ,
~\cos\left(2\gamma_3^{} - \alpha_2^{} -\alpha_3^{}\right) \; ;
\nonumber
\end{array}
\end{eqnarray}
\begin{eqnarray}
\begin{array}{l}
\cos\left(2\alpha_1^{} + \alpha_2^{} + \alpha_3^{}\right) \; ,
~\cos\left(2\alpha_2^{}+ \alpha_1^{} + \alpha_3^{}\right) \; ,
~\cos\left(2\alpha_3^{}+ \alpha_1^{} + \alpha_2^{}\right) \; ;
\\
\cos\left(2\beta_1^{} + \beta_2^{} + \beta_3^{}\right) \; ,
~\cos\left(2\beta_2^{}+ \beta_1^{} + \beta_3^{}\right) \; ,
~\cos\left(2\beta_3^{}+ \beta_1^{} + \beta_2^{}\right) \; ;
\\
\cos\left(2\gamma_1^{} + \gamma_2^{} + \gamma_3^{}\right) \; ,
~\cos\left(2\gamma_2^{}+ \gamma_1^{} + \gamma_3^{}\right) \; ,
~\cos\left(2\gamma_3^{}+ \gamma_1^{} + \gamma_2^{}\right) \; ;
\nonumber \\
\cos\left(\alpha_1^{} + \beta_2^{} +  \gamma_3^{}\right) \; ,
~\cos\left(\alpha_1^{} + \beta_3^{} +  \gamma_2^{}\right) \; ,
~\cos\left(\alpha_2^{} + \beta_1^{} +  \gamma_3^{}\right) \; ,
\\
\cos\left(\alpha_2^{} + \beta_3^{} +  \gamma_1^{}\right) \; ,
~\cos\left(\alpha_3^{} + \beta_1^{} +  \gamma_2^{}\right) \; ,
~\cos\left(\alpha_3^{} + \beta_2^{} +  \gamma_1^{}\right) \; ;
\nonumber \\	
\cos\left(\alpha_1^{} - \beta_1^{} + \gamma_2^{}\right) \; ,
~\cos\left(\alpha_1^{} - \beta_1^{} + \gamma_3^{}\right) \; ,
~\cos\left(\alpha_1^{} - \beta_2^{} + \gamma_2^{}\right) \; ,
~\cos\left(\alpha_1^{} - \beta_2^{} + \gamma_3^{}\right) \; ,
\\
\cos\left(\alpha_1^{} - \beta_3^{} + \gamma_2^{}\right) \; ,
~\cos\left(\alpha_1^{} - \beta_3^{} + \gamma_3^{}\right) \; ,
~\cos\left(\alpha_2^{} - \beta_1^{} + \gamma_1^{}\right) \; ,
~\cos\left(\alpha_2^{} - \beta_1^{} + \gamma_3^{}\right) \; ,
\\
\cos\left(\alpha_2^{} - \beta_2^{} + \gamma_1^{}\right) \; ,
~\cos\left(\alpha_2^{} - \beta_2^{} + \gamma_3^{}\right) \; ,
~\cos\left(\alpha_2^{} - \beta_3^{} + \gamma_1^{}\right) \; ,
~\cos\left(\alpha_2^{} - \beta_3^{} + \gamma_3^{}\right) \; ,
\\
\cos\left(\alpha_3^{} - \beta_1^{} + \gamma_1^{}\right) \; ,
~\cos\left(\alpha_3^{} - \beta_1^{} + \gamma_2^{}\right) \; ,
~\cos\left(\alpha_3^{} - \beta_2^{} + \gamma_1^{}\right) \; ,
~\cos\left(\alpha_3^{} - \beta_2^{} + \gamma_2^{}\right) \; ,
\\
\cos\left(\alpha_3^{} - \beta_3^{} + \gamma_1^{}\right) \; ,
~\cos\left(\alpha_3^{} - \beta_3^{} + \gamma_2^{}\right) \; ;
~\cos\left(\alpha_1^{} +\beta_2^{} - \gamma_1^{}\right) \; ,
~\cos\left(\alpha_1^{} + \beta_3^{} - \gamma_1^{}\right) \; ,
\\
\cos\left(\alpha_1^{} +\beta_2^{} - \gamma_2^{}\right) \; ,
~\cos\left(\alpha_1^{} + \beta_3^{} - \gamma_2^{}\right) \; ,
~\cos\left(\alpha_1^{} +\beta_2^{} - \gamma_3^{}\right) \; ,
~\cos\left(\alpha_1^{} + \beta_3^{} - \gamma_3^{}\right) \; ,
\\
\cos\left(\alpha_2^{} +\beta_1^{} - \gamma_1^{}\right) \; ,
~\cos\left(\alpha_2^{} + \beta_3^{} - \gamma_1^{}\right) \; ,
~\cos\left(\alpha_2^{} + \beta_1^{} - \gamma_2^{}\right) \; ,
~\cos\left(\alpha_2^{} + \beta_3^{} - \gamma_2^{}\right) \; ,
\\
\cos\left(\alpha_2^{} + \beta_1^{} - \gamma_3^{}\right) \; ,
~\cos\left(\alpha_2^{} + \beta_3^{} - \gamma_3^{}\right) \; ,
~\cos\left(\alpha_3^{} + \beta_1^{} - \gamma_1^{}\right) \; ,
~\cos\left(\alpha_3^{} + \beta_2^{} - \gamma_1^{}\right) \;,
\\
\cos\left(\alpha_3^{} + \beta_1^{} - \gamma_2^{}\right) \; ,
~\cos\left(\alpha_3^{} + \beta_2^{} - \gamma_2^{}\right) \; ,
~\cos\left(\alpha_3^{} + \beta_1^{} - \gamma_3^{}\right) \; ,
~\cos\left(\alpha_3^{} + \beta_2^{} - \gamma_3^{}\right) \; ;
\\
\cos\left(\beta_1^{} + \gamma_2^{} - \alpha_1^{}\right) \; ,
~\cos\left(\beta_1^{} + \gamma_3^{} - \alpha_1^{}\right) \; ,
~\cos\left(\beta_1^{} + \gamma_2^{} - \alpha_2^{}\right) \; ,
~\cos\left(\beta_1^{} + \gamma_3^{} - \alpha_2^{}\right) \; ,
\\
\cos\left(\beta_1^{} + \gamma_2^{} - \alpha_3^{}\right) \; ,
~\cos\left(\beta_1^{} + \gamma_3^{} - \alpha_3^{}\right) \; ,
~\cos\left(\beta_2^{} + \gamma_1^{} - \alpha_1^{}\right) \; ,
~\cos\left(\beta_2^{} + \gamma_3^{} - \alpha_1^{}\right) \; ,
\\
\cos\left(\beta_2^{} + \gamma_1^{} - \alpha_2^{}\right) \; ,
~\cos\left(\beta_2^{} + \gamma_3^{} - \alpha_2^{}\right) \; ,
~\cos\left(\beta_2^{} + \gamma_1^{} - \alpha_3^{}\right) \; ,
~\cos\left(\beta_2^{} + \gamma_3^{} - \alpha_3^{}\right) \; ,
\\
\cos\left(\beta_3^{} + \gamma_1^{} - \alpha_1^{}\right) \; ,
~\cos\left(\beta_3^{} + \gamma_2^{} - \alpha_1^{}\right) \; ,
~\cos\left(\beta_3^{} + \gamma_1^{} - \alpha_2^{}\right) \; ,
~\cos\left(\beta_3^{} + \gamma_2^{} - \alpha_2^{}\right) \; ,
\\
\cos\left(\beta_3^{} + \gamma_1^{} - \alpha_3^{}\right) \; ,
~\cos\left(\beta_3^{} + \alpha_2^{} - \alpha_3^{}\right) \; ;
\nonumber \\
\cos\left(\alpha_1^{} + \alpha_2^{} -\beta_1^{} - \beta_2^{}\right) \; ,
~\cos\left(\alpha_1^{} + \alpha_2^{} -\beta_2^{} - \beta_3^{}\right) \; ,
~\cos\left(\alpha_1^{} + \alpha_2^{} -\beta_3^{} - \beta_1^{}\right) \; ,
\\
\cos\left(\alpha_2^{} + \alpha_3^{} -\beta_1^{} - \beta_2^{}\right) \; ,
~\cos\left(\alpha_2^{} + \alpha_3^{} -\beta_2^{} - \beta_3^{}\right) \; ,
~\cos\left(\alpha_2^{} + \alpha_3^{} -\beta_3^{} - \beta_1^{}\right) \; ,
\\
\cos\left(\alpha_3^{} + \alpha_1^{} -\beta_1^{} - \beta_2^{}\right) \; ,
~\cos\left(\alpha_3^{} + \alpha_1^{} -\beta_2^{} - \beta_3^{}\right) \; ,
~\cos\left(\alpha_3^{} + \alpha_1^{} -\beta_3^{} - \beta_1^{}\right) \; ;
\\	
\cos\left(\beta_1^{} + \beta_2^{}-\gamma_1^{} - \gamma_2^{}\right) \; ,
~\cos\left(\beta_1^{} +\beta_2^{}-\gamma_2^{} -\gamma_3^{}\right) \; ,
~\cos\left(\beta_1^{} + \beta_2^{}-\gamma_3^{} - \gamma_1^{}\right) \; ,
\\
\cos\left(\beta_2^{} +\beta_3^{}-\gamma_1^{} - \gamma_2^{}\right) \; ,
~\cos\left(\beta_2^{} +\beta_3^{}-\gamma_2^{} - \gamma_3^{}\right) \; ,
~\cos\left(\beta_2^{} + \beta_3^{}-\gamma_3^{} - \gamma_1^{}\right) \; ,
\\
\cos\left(\beta_3^{} + \beta_1^{}-\gamma_1^{} - \gamma_2^{}\right) \; ,
~\cos\left(\beta_3^{} + \beta_1^{}-\gamma_2^{} - \gamma_3^{}\right) \; ,
~\cos\left(\beta_3^{} + \beta_1^{}-\gamma_3^{} -\gamma_1^{}\right) \; ;
\\
\cos\left(\gamma_1^{} + \gamma_2^{} -\alpha_1^{} - \alpha_2^{}\right) \; ,
~\cos\left(\gamma_1^{} + \gamma_2^{} -\alpha_2^{} - \alpha_3^{}\right) \; ,
~\cos\left(\gamma_1^{} + \gamma_2^{} -\alpha_3^{} - \alpha_1^{}\right) \; ,
\\
\cos\left(\gamma_2^{} + \gamma_3^{} -\alpha_1^{} - \alpha_2^{}\right) \; ,
~\cos\left(\gamma_2^{} + \gamma_3^{} -\alpha_2^{} - \alpha_3^{}\right) \; ,
~\cos\left(\gamma_2^{} + \gamma_3^{} -\alpha_3^{} - \alpha_1^{}\right) \; ,
\\
\cos\left(\gamma_3^{} + \gamma_1^{} -\alpha_1^{} - \alpha_2^{}\right) \; ,
~\cos\left(\gamma_3^{} + \gamma_1^{} -\alpha_2^{} - \alpha_3^{}\right) \; ,
~\cos\left(\gamma_3^{} + \gamma_1^{} -\alpha_3^{} - \alpha_1^{}\right) \; .
\nonumber
\end{array}
\end{eqnarray}
In comparison, ${\cal J}_{\nu}^{}$ depends on 240 different combinations
of the original CP-violating phases in the canonical seesaw mechanism,
as explicitly shown in Ref.~\cite{Xing:2024xwb}.



\begin{thebibliography}{99}

\bibitem{Minkowski:1977sc}
P.~Minkowski,
``$\mu \to e\gamma$ at a rate of one out of $10^{9}$ muon decays?,''
Phys.\ Lett.\  {\bf 67B} (1977) 421.

\bibitem{Yanagida:1979as}
T.~Yanagida,
``Horizontal gauge symmetry and masses of neutrinos,''
Conf.\ Proc.\ C {\bf 7902131} (1979) 95.

\bibitem{GellMann:1980vs}
M.~Gell-Mann, P.~Ramond and R.~Slansky,
``Complex spinors and unified theories,''
Conf.\ Proc.\ C {\bf 790927} (1979) 315
[arXiv:1306.4669 [hep-th]].

\bibitem{Glashow:1979nm}
S.~L.~Glashow,
``The future of elementary particle physics,''
NATO Sci.\ Ser.\ B {\bf 61} (1980) 687.

\bibitem{Mohapatra:1979ia}
R.~N.~Mohapatra and G.~Senjanovic,
``Neutrino mass and spontaneous parity nonconservation,''
Phys.\ Rev.\ Lett.\  {\bf 44} (1980) 912.

\bibitem{Fukugita:1986hr}
M.~Fukugita and T.~Yanagida,
``Baryogenesis Without Grand Unification,''
Phys. Lett. B \textbf{174} (1986), 45-47.

\bibitem{Xing:2007zj}
Z.~z.~Xing,
``Correlation between the Charged Current Interactions of Light and Heavy Majorana
Neutrinos,''
Phys. Lett. B \textbf{660} (2008), 515-521
[arXiv:0709.2220 [hep-ph]].

\bibitem{Xing:2011ur}
Z.~z.~Xing,
``A full parametrization of the 6$\times$6 flavor mixing matrix in the presence of
three light or heavy sterile neutrinos,''
Phys. Rev. D \textbf{85} (2012), 013008
[arXiv:1110.0083 [hep-ph]].

\bibitem{Pontecorvo:1957cp}
B.~Pontecorvo,
``Mesonium and anti-mesonium,''
Sov.\ Phys.\ JETP {\bf 6} (1957) 429
[Zh.\ Eksp.\ Teor.\ Fiz.\  {\bf 33} (1957) 549].

\bibitem{Maki:1962mu}
Z.~Maki, M.~Nakagawa and S.~Sakata,
``Remarks on the unified model of elementary particles,''
Prog.\ Theor.\ Phys.\  {\bf 28} (1962) 870.

\bibitem{Pontecorvo:1967fh}
B.~Pontecorvo,
``Neutrino Experiments and the Problem of Conservation of Leptonic Charge,''
Sov.\ Phys.\ JETP {\bf 26} (1968) 984
[Zh.\ Eksp.\ Teor.\ Fiz.\  {\bf 53} (1967) 1717].

\bibitem{JUNO:2015zny}
F.~An \textit{et al.} [JUNO],
``Neutrino Physics with JUNO,''
J. Phys. G \textbf{43} (2016) no.3, 030401
[arXiv:1507.05613 [physics.ins-det]].

\bibitem{JUNO:2021vlw}
A.~Abusleme \textit{et al.} [JUNO],
``JUNO physics and detector,''
Prog. Part. Nucl. Phys. \textbf{123} (2022), 103927
[arXiv:2104.02565 [hep-ex]].

\bibitem{DUNE:2015lol}
R.~Acciarri \textit{et al.} [DUNE],
``Long-Baseline Neutrino Facility (LBNF) and Deep Underground Neutrino Experiment (DUNE): Conceptual Design Report, Volume 2: The Physics Program for DUNE at LBNF,''
[arXiv:1512.06148 [physics.ins-det]].

\bibitem{Hyper-KamiokandeProto-:2015xww}
K.~Abe \textit{et al.} [Hyper-Kamiokande Proto-],
``Physics potential of a long-baseline neutrino oscillation experiment using a J-PARC
neutrino beam and Hyper-Kamiokande,''
PTEP \textbf{2015} (2015), 053C02
[arXiv:1502.05199 [hep-ex]].

\bibitem{Xing:2023adc}
Z.~z.~Xing,
``The formal seesaw mechanism of Majorana neutrinos with unbroken gauge symmetry,''
Nucl. Phys. B \textbf{987} (2023), 116106
[arXiv:2301.10461 [hep-ph]].

\bibitem{Weinberg:1979sa}
S.~Weinberg,
``Baryon and Lepton Nonconserving Processes,''
Phys. Rev. Lett. \textbf{43} (1979), 1566-1570

\bibitem{ParticleDataGroup:2024cfk}
S.~Navas \textit{et al.} [Particle Data Group],
``Review of particle physics,''
Phys. Rev. D \textbf{110} (2024) no.3, 030001

\bibitem{Jarlskog:1985ht}
C.~Jarlskog,
``Commutator of the Quark Mass Matrices in the Standard Electroweak Model and a Measure of Maximal $CP$~Nonconservation,''
Phys. Rev. Lett. \textbf{55} (1985), 1039.

\bibitem{Wu:1985ea}
D.~d.~Wu,
``The Rephasing Invariants and CP,''
Phys. Rev. D \textbf{33} (1986), 860.

\bibitem{Xing:2024xwb}
Z.~z.~Xing,
``Mapping the sources of CP violation in neutrino oscillations from the seesaw
mechanism,''
Phys. Lett. B \textbf{856} (2024), 138909
[arXiv:2406.01142 [hep-ph]].

\bibitem{Xing:2020ijf}
Z.~z.~Xing,
``Flavor structures of charged fermions and massive neutrinos,''
Phys. Rept. \textbf{854} (2020), 1-147
[arXiv:1909.09610 [hep-ph]].

\bibitem{Antusch:2014woa}
S.~Antusch and O.~Fischer,
``Non-unitarity of the leptonic mixing matrix: Present bounds and future sensitivities,''
JHEP \textbf{10} (2014), 094
[arXiv:1407.6607 [hep-ph]].

\bibitem{Blennow:2016jkn}
M.~Blennow, P.~Coloma, E.~Fernandez-Martinez, J.~Hernandez-Garcia and J.~Lopez-Pavon,
``Non-Unitarity, sterile neutrinos, and Non-Standard neutrino Interactions,''
JHEP \textbf{04} (2017), 153
[arXiv:1609.08637 [hep-ph]].

\bibitem{Wang:2021rsi}
Y.~Wang and S.~Zhou,
``Non-unitary leptonic flavor mixing and CP violation in neutrino-antineutrino
oscillations,''
Phys. Lett. B \textbf{824} (2022), 136797
[arXiv:2109.13622 [hep-ph]].

\bibitem{Blennow:2023mqx}
M.~Blennow, E.~Fern\'andez-Mart\'\i{}nez, J.~Hern\'andez-Garc\'\i{}a, J.~L\'opez-Pav\'on, X.~Marcano and D.~Naredo-Tuero,
``Bounds on lepton non-unitarity and heavy neutrino mixing,''
JHEP \textbf{08} (2023), 030
[arXiv:2306.01040 [hep-ph]].

\bibitem{Planck:2018vyg}
N.~Aghanim \textit{et al.} [Planck],
``Planck 2018 results. VI. Cosmological parameters,''
Astron. Astrophys. \textbf{641} (2020), A6
[erratum: Astron. Astrophys. \textbf{652} (2021), C4]
[arXiv:1807.06209 [astro-ph.CO]].

\bibitem{Fengler:2024ufb}
C.~Fengler [KATRIN],
``A Look at General Neutrino Interactions with KATRIN,''
PoS \textbf{EPS-HEP2023} (2024), 188

\bibitem{Agostini:2022zub}
M.~Agostini, G.~Benato, J.~A.~Detwiler, J.~Men\'endez and F.~Vissani,
``Toward the discovery of matter creation with neutrinoless \ensuremath{\beta}\ensuremath{\beta} decay,''
Rev. Mod. Phys. \textbf{95} (2023) no.2, 025002
[arXiv:2202.01787 [hep-ex]].

\bibitem{Xing:2020ivm}
Z.~z.~Xing and D.~Zhang,
``Radiative decays of charged leptons as constraints of unitarity polygons for
active-sterile neutrino mixing and CP violation,''
Eur. Phys. J. C \textbf{80} (2020) no.12, 1134
[arXiv:2009.09717 [hep-ph]].

\bibitem{Kleppe:1995zz}
A.~Kleppe,
``Extending the standard model with two right-handed neutrinos,''
in {\it Neutrino physics} (Proceedings of the 3rd Tallinn Symposium, Lohusalu, Estonia,
October 8-11, 1995), page 118-125.

\bibitem{Ma:1998zg}
E.~Ma, D.~P.~Roy and U.~Sarkar,
``A Seesaw model for atmospheric and solar neutrino oscillations,''
Phys. Lett. B \textbf{444} (1998), 391-396
[arXiv:hep-ph/9810309 [hep-ph]].

\bibitem{Frampton:2002qc}
P.~H.~Frampton, S.~L.~Glashow and T.~Yanagida,
``Cosmological sign of neutrino CP violation,''
Phys. Lett. B \textbf{548} (2002), 119-121
[arXiv:hep-ph/0208157 [hep-ph]].

\bibitem{Endoh:2002wm}
T.~Endoh, S.~Kaneko, S.~K.~Kang, T.~Morozumi and M.~Tanimoto,
``CP violation in neutrino oscillation and leptogenesis,''
Phys. Rev. Lett. \textbf{89} (2002), 231601
[arXiv:hep-ph/0209020 [hep-ph]].

\bibitem{Raidal:2002xf}
M.~Raidal and A.~Strumia,
``Predictions of the most minimal seesaw model,''
Phys. Lett. B \textbf{553} (2003), 72-78
[arXiv:hep-ph/0210021 [hep-ph]].

\bibitem{Dutta:2003ps}
B.~Dutta and R.~N.~Mohapatra,
``Lepton flavor violation and neutrino mixings in a 3 x 2 seesaw model,''
Phys. Rev. D \textbf{68} (2003), 056006
[arXiv:hep-ph/0305059 [hep-ph]].

\bibitem{Ibarra:2003xp}
A.~Ibarra and G.~G.~Ross,
``Neutrino properties from Yukawa structure,''
Phys. Lett. B \textbf{575} (2003), 279-289
[arXiv:hep-ph/0307051 [hep-ph]].

\bibitem{Xing:2020ald}
Z.~z.~Xing and Z.~h.~Zhao,
``The minimal seesaw and leptogenesis models,''
Rept. Prog. Phys. \textbf{84} (2021) no.6, 066201
[arXiv:2008.12090 [hep-ph]].

\bibitem{Xing:2023kdj}
Z.~z.~Xing,
``CP violation in light neutrino oscillations and heavy neutrino decays: A general and explicit
seesaw-bridged correlation,''
Phys. Lett. B \textbf{844} (2023), 138065
[arXiv:2306.02362 [hep-ph]].

\bibitem{Ohlsson:2013xva}
T.~Ohlsson and S.~Zhou,
``Renormalization group running of neutrino parameters,''
Nature Commun. \textbf{5} (2014), 5153
[arXiv:1311.3846 [hep-ph]].

\end{thebibliography}
\end{document}